\documentclass[pdflatex,sn-basic]{sn-jnl}

\usepackage{graphicx}%
\usepackage{multirow}%
\usepackage{amsmath,amssymb,amsfonts}%
\usepackage{amsthm}%
\usepackage{mathrsfs}%
\usepackage[title]{appendix}%
\usepackage{xcolor}%
\usepackage{textcomp}%
\usepackage{manyfoot}%
\usepackage{booktabs}%
\usepackage{algorithm}%
\usepackage{algorithmicx}%
\usepackage{algpseudocode}%
\usepackage{listings}%
\usepackage{natbib}
\usepackage{pdflscape}

\usepackage{subcaption}
\usepackage{enumitem}

\theoremstyle{thmstyleone}%
\theoremstyle{thmstyletwo}%

\theoremstyle{thmstylethree}%

\begin{document}

\title[Energy Balancing Weights for Mediation Analysis]{Energy Balancing Weights for Mediation Analysis}


\author*[1,2]{\fnm{Taishi} \sur{Odaka}}\email{gds9525503@juntendo.ac.jp}
\author[1]{\fnm{Kentaro} \sur{Sakamaki}}\email{kentaro.sakamaki@gmail.com}

\affil*[1]{\orgdiv{Graduate School of Health Data Science}, \orgname{Juntendo University}, \orgaddress{\street{Hinode}, \city{Urayasu}, \postcode{279-0013}, \state{Chiba}, \country{Japan}}}

\affil[2]{\orgdiv{Biostatistics, Asia Research and Development}, \orgname{GlaxoSmithKline K.K.}, \orgaddress{\street{Akasaka}, \city{Minato-ku}, \postcode{107-0052}, \state{Tokyo}, \country{Japan}}}


\abstract{
Causal mediation analysis requires reconstruction of counterfactual distributions to estimate natural direct and indirect effects. Inverse probability weighting estimators rely on models for treatment assignment and mediator density ratios, whereas moment balancing approaches require researchers to specify in advance which functions of the covariates and mediators should be balanced.

We propose Energy Balancing Weights for Mediation Analysis (EBWMA), which targets the joint mediator--covariate distribution used to identify counterfactual means such as \(E[Y(1,M(0))]\). Under standard identification conditions for natural effects, EBWMA constructs weights whose weighted empirical distribution approximates this target, without modeling treatment assignment, mediator density ratios, or the outcome regression. The weights minimize energy distance through two quadratic programming problems solved sequentially.

In simulations with nonlinearly transformed, skewed, or binary covariates and nonlinear mediator and outcome models, EBWMA generally achieved favorable bias and root mean squared error, with uniformly lower Monte Carlo variability than gradient boosting-based inverse probability weighting and moment balancing weights. Under the nonlinearly transformed structure, gradient boosting performed better for the natural indirect effect, although the gap narrowed markedly under an alternative outcome model. This indicates sensitivity to the outcome regression rather than to covariate geometry alone. In an illustrative analysis of the National Health and Nutrition Examination Survey I Epidemiologic Follow-up Study, EBWMA gave the smallest standardized mean differences for most covariates and for the mediator.

EBWMA requires no outcome model and no balance functions or tolerance parameters, although preprocessing choices remain consequential. Because energy distance reflects the metric geometry of the mediator--covariate space, applications should assess weight stability and distributional balance.
}


\maketitle


\section{Introduction}
\label{sec1}

Causal mediation analysis aims to decompose the total effect of a treatment into components operating through and not through a specified mediator. The central estimands are the natural direct effect (NDE) and natural indirect effect (NIE), defined through counterfactual means of the form $\mu_{a,a'} = E[Y(a, M(a'))]$: the expected outcome when the treatment is set to $a$ while the mediator follows the distribution it would have had under treatment $a'$. For $a \neq a'$, these are cross-world quantities. Whereas average treatment effect (ATE) estimation requires reconstructing only the marginal covariate distribution, estimating the NDE and NIE requires reconstructing the joint distribution of $(M, X)$ of the form $dF_{M \mid A=a', X=x}(m)\,dF_X(x)$, since the mediator distribution is defined conditional on baseline covariates $X$. This joint distribution generally does not coincide with the observed joint distribution in either treatment arm, and reconstructing it is challenging in practice, particularly when the mediator and covariates are related through nonlinear or interacting mechanisms or when $X$ is high-dimensional.

Weighting-based estimators provide one way to implement this reconstruction. Inverse probability weighting (IPW) estimators for mediation analysis combine weights for treatment assignment with density-ratio weights for the mediator distribution, thereby using outcomes from individuals observed under the treatment level $a$ while reweighting their mediator and covariate distribution toward the one induced by the treatment level $a'$ \citep{huber2014identifying,hong2025ratio}. This strategy avoids specifying the outcome regression, but shifts the modeling burden to the treatment assignment and mediator distribution models. Because these components enter the weights multiplicatively, misspecification of either can distort the reconstructed target distribution and produce unstable weights, particularly when the treatment, mediator, and covariates are related through nonlinear or interacting mechanisms \citep{chattopadhyay2024one}. Semiparametric efficient estimators \citep{tchetgen2012semiparametric,zheng2012targeted} mitigate this sensitivity by augmenting IPW-type weights with an outcome regression component, which confers a multiple robustness property: consistency is retained when a subset of the component models is correctly specified. Their performance can nonetheless degrade when all component models are misspecified, and specifying each model remains a practical burden.

An alternative strategy is to obtain weights by imposing balance conditions directly, rather than by specifying treatment, mediator, or outcome models. Moment balancing methods formulate the desired balance as constraints in an optimization problem and have been developed for both average treatment effect estimation and causal mediation analysis \citep{hainmueller2012entropy,zubizarreta2015stable,chan2016globally,chan2016efficient,kawato2025balancing}. In the mediation setting, such methods balance prespecified functions \(v_k(M,X)\) of the mediator and covariates, which may include main effects, nonlinear transformations, and interaction terms: the weights are chosen so that the weighted empirical moments of these functions among individuals with \(A=a\) approximate the corresponding moments under the target joint distribution of \((M,X)\). This replaces model specification with the selection of balance functions, and bias reduction is most transparent when the chosen functions capture features of \((M,X)\) that are relevant to the outcome regression. In practice, however, it is rarely clear which transformations or interactions of \((M,X)\) matter, and omitted features may leave discrepancies that are consequential for the counterfactual mean. This difficulty is more acute than in the ATE setting, since the relevant functions involve the joint behavior of the mediator and covariates rather than the covariates alone.

Energy Balancing Weights (EBW), proposed by \cite{huling2024energy}, avoid prespecifying a finite set of balance functions by minimizing the energy distance \citep{szekely2013energy,rizzo2016energy} between a weighted empirical distribution and a target empirical distribution. For distributions with finite first moments, the energy distance is nonnegative and equals zero if and only if the two distributions coincide \citep{szekely2013energy}. Minimizing it therefore targets discrepancies between entire distributions rather than discrepancies in finitely many prespecified moments. In the ATE setting, EBW targets the marginal covariate distribution \(F_X\) and has been shown to yield stable weights with favorable finite-sample performance \citep{huling2024energy}.

Extending EBW to causal mediation analysis, however, is not immediate. The relevant target is no longer \(F_X\) but the joint distribution
\[
dF_{a'}^\star(m,x) = dF_{M \mid A=a',X=x}(m)\,dF_X(x),
\]
which combines the conditional mediator distribution under one treatment level with the marginal covariate distribution of the target population. In general, neither arm directly represents \(F_{a'}^\star\): among individuals with \(A=a'\), the conditional mediator distribution is the required one, but the accompanying covariate distribution is \(F_{X\mid A=a'}\) rather than \(F_X\). This observation motivates a two-stage energy-balancing construction, in which first-stage weights targeting \(F_X\) define the empirical target for second-stage balancing of the joint distribution of \((M,X)\).

In this study, we propose Energy Balancing Weights for Mediation Analysis (EBWMA), a weighting method that extends energy-distance-based distribution balancing to causal mediation analysis. The central idea is to recast weighting-based mediation analysis as the empirical reconstruction of the target joint distribution \(F_{a'}^\star\) of \(V=(M,X)\). This formulation makes explicit the distributional target that conventional IPW and density-ratio estimators recover only implicitly, through model-based weighting components. In doing so, it points to an alternative estimation strategy: EBWMA reconstructs \(F_{a'}^\star\) through a two-stage energy-distance minimization procedure, requiring neither treatment assignment models nor mediator density-ratio models nor a prespecified finite set of balance functions. Natural direct and indirect effects are then estimated from weighted outcome means. We assess the finite-sample performance and diagnostic behavior of EBWMA through simulations and an empirical analysis of the NHEFS data, with particular emphasis on distributional balance, weight stability, and sensitivity to skewed or heavy-tailed data structures.

The remainder of this paper is organized as follows. Section~\ref{sec2} reviews the mediation identification framework and weighting-based estimation. Section~\ref{sec3} summarizes moment balancing and energy balancing weights. Section~\ref{sec4} introduces EBWMA. Sections~\ref{sec5} and~\ref{sec6} present the simulation and empirical results, respectively. Section~\ref{sec7} concludes with limitations and future directions.


\section{Definition and Identification}
\label{sec2}

\subsection{Potential Outcomes and the ATE Framework}

Let $O_i=(X_i,A_i,M_i,Y_i)$, $i=1,\ldots,n$, be independent and identically distributed (i.i.d.) copies of $O=(X,A,M,Y)$, where $A\in\{0,1\}$ is a binary treatment, $X\in\mathcal{X}\subset\mathbb{R}^p$ is a vector of pre-treatment covariates, $M\in\mathcal{M}\subset\mathbb{R}$ is a mediator, and $Y\in\mathcal{Y}\subset\mathbb{R}$ is an outcome. Throughout, $a,a'\in\{0,1\}$, and $\mathcal{I}_a=\{i:A_i=a\}$ denotes the index set of individuals with $A_i=a$, with $n_a=|\mathcal{I}_a|$.

Under the potential outcomes framework, each individual has a potential outcome $Y(a)$ under treatment $a$, and the ATE is defined as
\[
\tau_\mathrm{ATE}=E\left[Y(1)-Y(0)\right].
\]
The ATE is identified from the observed data under the following conditions:
\begin{itemize}
\item[(A1)] Consistency: $Y=AY(1)+(1-A)Y(0)$.
\item[(A2)] Exchangeability: $\{Y(0),Y(1)\}\perp A\mid X$.
\item[(A3)] Positivity: $0<\Pr(A=1\mid X=x)<1$ for all $x\in\mathcal{X}$ with $dF_X(x)>0$.
\end{itemize}
Under (A1)--(A3),
\[
E\left[Y(a)\right]=E_X\left[E\left[Y\mid A=a,X\right]\right],
\]
where the outer expectation is taken with respect to the marginal distribution $F_X$.

\subsection{Causal Mediation Framework and Identification Conditions}

Causal mediation analysis decomposes the effect of a treatment into components that act through a mediator and components that do not. Let $M(a)$ denote the potential mediator under treatment $a$, and let $Y(a,m)$ denote the potential outcome when the treatment is set to $a$ and the mediator to $m$. We assume the composition relation $Y(a)=Y(a,M(a))$.

We focus on counterfactual means of the form
\[
\mu_{a,a'}=E\left[Y\left(a,M(a')\right)\right],
\]
where the first subscript $a$ is the treatment value entering the outcome and the second subscript $a'$ is the treatment value generating the mediator distribution. The total effect (TE), natural direct effect (NDE), and natural indirect effect (NIE) are defined as
\begin{align}
\tau_\mathrm{TE} &=\mu_{1,1}-\mu_{0,0},\\
\tau_\mathrm{NDE}&=\mu_{1,0}-\mu_{0,0},\\
\tau_\mathrm{NIE}&=\mu_{1,1}-\mu_{1,0},
\end{align}
respectively, so that $\tau_\mathrm{TE}=\tau_\mathrm{NDE}+\tau_\mathrm{NIE}$. Note that $\tau_\mathrm{TE}$ coincides with the ATE, whereas $\mu_{1,0}$ involves the cross-world quantity $Y(1,M(0))$, which is not observed for any individual.

The counterfactual mean $\mu_{a,a'}$ is identified from the observed data under the following conditions, which hold for all $a,a'\in\{0,1\}$ and all $m\in\mathcal{M}$:
\begin{itemize}
\item[(B1)] Consistency: if $A=a$, then $M=M(a)$; if $A=a$ and $M=m$, then $Y=Y(a,m)$.
\item[(B2)] Treatment exchangeability: $\{Y(a,m),M(a')\}\perp A\mid X$.
\item[(B3)] Cross-world exchangeability: $Y(a,m)\perp M(a')\mid X$.
\item[(B4)] Positivity: $\Pr(A=a\mid X=x)>0$ for all $x\in\mathcal{X}$ with $dF_X(x)>0$, and the density ratio
\[
\frac{dF_{M\mid A=a',X=x}(m)}
{dF_{M\mid A=a,X=x}(m)}
\]
is well defined on the support relevant to the analysis.
\end{itemize}
Condition (B2), together with consistency, allows the distributions of the potential mediator and potential outcome under each treatment level to be identified from the corresponding observed treatment groups after conditioning on $X$. Condition (B3) is the cross-world assumption specific to natural effects. It states that, conditional on $X$, the potential outcome under $(a,m)$ is independent of the mediator that would arise under $a'$, and therefore permits the outcome response under one treatment world to be combined with the mediator distribution from another. This condition is stronger than ordinary adjustment for measured mediator--outcome confounding and cannot be verified from observed data, even in a randomized experiment. In particular, the standard mediation formula does not generally apply in the presence of mediator--outcome confounders that are themselves affected by treatment. Conditions (B1)--(B4) are therefore substantially stronger than (A1)--(A3).
Under (B1)--(B4), $\mu_{a,a'}$ is identified by the mediation formula \citep{valeri2013mediation,vanderweele2014unification,vanderweele2017mediation}:
\begin{equation}
\mu_{a,a'}
=\int E\left[Y\mid A=a,M=m,X=x\right] \, dF_{M\mid A=a',X=x}(m) \, dF_X(x).
\label{eq:mediation_formula}
\end{equation}
This expression combines the outcome regression under $A=a$ with the mediator distribution under $A=a'$ within levels of $X$, and then averages over the target covariate distribution $F_X$.

\subsection{Target Joint Distribution and Its Weighting Representation}

We now rewrite the identification formula to make explicit the distributional target of weighting estimators. Let
\[
g_a(m,x)=E\left[Y\mid A=a,M=m,X=x\right]
\]
denote the observed-data outcome regression in treatment group $A=a$, and, for $V=(M,X)$, define the target joint distribution $F_{a'}^\star$ by
\begin{equation}
dF_{a'}^\star(m,x)
=
dF_{M\mid A=a',X=x}(m)\,dF_X(x).
\label{eq:target_joint_distribution}
\end{equation}
The mediation formula \eqref{eq:mediation_formula} can then be written as
\begin{equation}
\mu_{a,a'}=\int g_a(m,x)\,dF_{a'}^\star(m,x).
\label{eq:mediation_formula_joint}
\end{equation}

Representation \eqref{eq:mediation_formula_joint} underlies the proposed method. Although the mediation formula involves the conditional mediator distribution $F_{M\mid A=a',X=x}$, the counterfactual mean depends on it only through the induced joint distribution $F_{a'}^\star$ of $V$. For $\mu_{1,0}$, in particular,
\[
\mu_{1,0}=\int g_1(m,x)\,dF_0^\star(m,x),
\qquad
dF_0^\star(m,x)=dF_{M\mid A=0,X=x}(m)\,dF_X(x).
\]
This observation motivates EBWMA, which reconstructs $F_{a'}^\star$ at the level of empirical distributions rather than estimating the conditional mediator density pointwise in $x$.

We next describe the conventional weighting representation of \eqref{eq:mediation_formula}. Let $\pi_a(x)=\Pr(A=a\mid X=x)$ denote the treatment assignment probability. For the ATE, the normalized IPW estimator
\[
\widehat{E}\left[Y(a)\right]=
\frac{\displaystyle\sum_{i\in\mathcal{I}_a}Y_i/\widehat{\pi}_a(X_i)}
{\displaystyle\sum_{i\in\mathcal{I}_a}1/\widehat{\pi}_a(X_i)}
\]
is conventionally used.

In mediation analysis, one must use outcomes from individuals with ($A=a$) while reweighting their joint mediator--covariate distribution toward the target that combines the mediator distribution under ($A=a'$) with the marginal covariate distribution ($F_X$). Following \cite{huber2014identifying} and \cite{hong2025ratio}, this operation is implemented through the ratio weight
\begin{equation}
\begin{aligned}
w_i^{a,a'}
&=
\frac{dF_{M\mid A=a',X=x_i}(m_i)}
     {dF_{M\mid A=a,X=x_i}(m_i)}
\cdot\frac{1}{\pi_a(x_i)} \\[2pt]
&=
\frac{\Pr(A=a'\mid M=m_i,X=x_i)}
     {\Pr(A=a\mid M=m_i,X=x_i)}
\cdot\frac{1}{\pi_{a'}(x_i)},
\qquad i\in\mathcal{I}_a,
\end{aligned}
\label{eq:ratio_weight_general}
\end{equation}
where the second expression follows from Bayes' theorem. This form is particularly convenient when the mediator is continuous, because it replaces the conditional mediator density with treatment probability models conditional on $(M,X)$; we therefore adopt it in the simulations and the real-data analysis. When $a=a'$, the density ratio equals one and $w_i^{a,a}$ reduces to the standard inverse probability weight. The corresponding normalized estimator is
\begin{equation}
\widehat{\mu}_{a,a'}
=
\frac{\displaystyle\sum_{i\in\mathcal{I}_a}\widehat{w}_i^{a,a'}Y_i}
{\displaystyle\sum_{i\in\mathcal{I}_a}\widehat{w}_i^{a,a'}}.
\label{eq:weighted_mu}
\end{equation}

Equation \eqref{eq:ratio_weight_general} shows that conventional IPW-type weights require a treatment assignment model together with either a conditional mediator density model or, equivalently, a treatment probability model conditional on $(M,X)$. These components enter the weights multiplicatively, so misspecification of either **can distort** the empirical reconstruction of $F_{a'}^\star$ and lead to bias in $\widehat{\mu}_{a,a'}$.

The method proposed in this study reduces this model dependence by balancing empirical distributions directly. EBWMA targets the same joint distribution $F_{a'}^\star$ required by the mediation formula, but specifies neither the treatment assignment model nor the conditional mediator density model. This does not relax the identification conditions (B1)--(B4); it provides an alternative estimation strategy once \eqref{eq:mediation_formula} is assumed to hold. In the next section, we review moment balancing and energy balancing as tools for constructing such empirical target distributions.


\section{Moment and Energy Balancing}
\label{sec3}

\subsection{Moment Balancing in the ATE Setting}

Rather than estimating weighting functions through explicit models, moment balancing methods construct weights by imposing balance conditions directly through an optimization problem \citep{hainmueller2012entropy,zubizarreta2015stable,chan2016globally}. These methods can be viewed within the minimal weights framework of \cite{wang2020minimal}, in which weights are chosen to make the weighted empirical distribution of a selected group resemble a target distribution with respect to prespecified functions of the covariates.

Let
\[
u(X)=\left(u_1(X),\ldots,u_K(X)\right)^\top
\]
denote a vector of prespecified basis functions of the covariates. For estimating $E[Y(a)]$, the weights $w^{a,a}=(w_i^{a,a})_{i\in\mathcal{I}_a}$ for individuals in treatment group $A=a$ are obtained as the solution to
\begin{equation}
\begin{aligned}
\min_{w^{a,a}} \quad
&\sum_{i\in\mathcal{I}_a} h(w_i^{a,a}) \\
\text{s.t.}\quad
&\left|\sum_{i\in\mathcal{I}_a} w_i^{a,a} u_k(X_i)
-\frac{1}{n}\sum_{i=1}^{n}u_k(X_i)\right|
\leq\delta_k,
\quad k=1,\ldots,K,\\
&\sum_{i\in\mathcal{I}_a} w_i^{a,a}=1,\quad w_i^{a,a}\geq 0,\quad i\in\mathcal{I}_a .
\end{aligned}
\label{eq:mw_ate}
\end{equation}
Here $h$ is a convex dispersion function that penalizes departures from uniform weights, such as $h(w)=w\log w$ or $h(w)=w^2/2$, and $\delta_k\geq 0$ is the tolerance for the $k$th balance constraint. The constraints require the weighted moments of $u_k$ among individuals with $A=a$ to lie within $\delta_k$ of the corresponding sample moments over the full sample, which serve as the empirical counterpart of the target marginal distribution $F_X$.

Problem \eqref{eq:mw_ate} thus approximates $F_X$ by constraining discrepancies in finitely many empirical moments. If the functions $u_k$ capture features of the covariate distribution that are relevant to the outcome regression, the resulting weighted mean can reduce confounding bias. The choice of $u_k$, however, is left to the researcher, and bias may remain if important nonlinear transformations or interactions are omitted from the balancing functions.

\subsection{Moment Balancing for the Target Joint Distribution in Mediation Analysis}

In mediation analysis, the distributional target is more complex than in ATE estimation. As shown in Section~\ref{sec2}, the counterfactual mean $\mu_{a,a'}=\int g_a(m,x)\,dF_{a'}^\star(m,x)$ depends on the joint distribution $F_{a'}^\star$ of $V=(M,X)$, so balancing the marginal distribution of $X$ alone does not suffice for estimands such as $\mu_{1,0}$. One must instead construct a weighted empirical distribution of $V$ that approximates $F_0^\star$.

Let
\[
v(M,X)=\left(v_1(M,X),\ldots,v_K(M,X)\right)^\top
\]
denote a vector of prespecified basis functions of $V$. Suppose the control-group weights $w^{0,0}$, corresponding to $\mu_{0,0}$, have already been obtained by balancing the marginal distribution of $X$ as in \eqref{eq:mw_ate}. Because the first-stage weighting uses only $X$, it is designed to shift the covariate distribution toward $F_X$ without altering the conditional mediator distribution given $A=0,X$. The induced weighted empirical distribution of $V$ among controls therefore serves as an empirical approximation to $F_0^\star$. Taking this distribution as the target, the treatment-group weights $w^{1,0}$ are obtained by solving
\begin{equation}
\begin{aligned}
\min_{w^{1,0}} \quad
&\sum_{i\in\mathcal{I}_1} h(w_i^{1,0}) \\
\text{s.t.}\quad
&\left|\sum_{i\in\mathcal{I}_1} w_i^{1,0} v_k(M_i,X_i)
-\sum_{i\in\mathcal{I}_0} w_i^{0,0} v_k(M_i,X_i)\right|
\leq\delta_k,
\quad k=1,\ldots,K,\\
&\sum_{i\in\mathcal{I}_1} w_i^{1,0}=1,\quad w_i^{1,0}\geq 0,\quad i\in\mathcal{I}_1 .
\end{aligned}
\label{eq:mw_mediation}
\end{equation}
Unlike in \eqref{eq:mw_ate}, the target moments are themselves estimated in the first stage through $w^{0,0}$ and are then treated as fixed in the second-stage optimization. Given $w^{1,0}$, the counterfactual mean $\mu_{1,0}$ is estimated by the weighted mean of $Y$ among individuals with $A=1$, analogously to \eqref{eq:weighted_mu}.

Setting $\delta_k=0$ yields exact moment balance, when feasible, as in \cite{chan2016efficient}, whereas $\delta_k>0$ gives the approximate balance considered by \cite{kawato2025balancing}. In both cases, $F_0^\star$ is approximated through finitely many constraints of the form
\[
\int v_k(m,x)\,d\widehat F_{1,w^{1,0}}^V(m,x)
\approx
\int v_k(m,x)\,d\widehat F_{0,w^{0,0}}^V(m,x),
\quad k=1,\ldots,K,
\]
where $\widehat F_{1,w^{1,0}}^V$ and $\widehat F_{0,w^{0,0}}^V$ denote the weighted empirical distributions of $V$ in the treatment and control groups, respectively.

This formulation exhibits both the strength and the limitation of moment balancing in mediation analysis. It requires no explicit specification of the treatment assignment model or the mediator density model, but it does require the researcher to prespecify which functions $v_k(M,X)$ to balance. That choice is harder than in the ATE setting, since the relevant features now concern the joint behavior of the mediator and covariates rather than the covariates alone. If transformations or interactions of $(M,X)$ that matter for the outcome regression $g_1(m,x)$ are omitted from $v(M,X)$, the resulting estimator of $\mu_{1,0}$ may remain biased. This motivates a distributional discrepancy criterion that does not require selecting basis functions at all.

\subsection{Energy Balancing Weights in the ATE Setting}

Energy Balancing Weights (EBW), proposed by \cite{huling2024energy}, provide such a criterion. Rather than matching finitely many moments, EBW minimizes the energy distance between a weighted empirical distribution and a target empirical distribution \citep{szekely2013energy,rizzo2016energy}, thereby replacing prespecified moment constraints with a general-purpose measure of distributional discrepancy.

Let $P$ and $Q$ be probability distributions on $\mathbb{R}^d$ with finite first moments, and let $\mathbf{S},\mathbf{S}'\overset{\mathrm{iid}}{\sim}P$ and $\mathbf{T},\mathbf{T}'\overset{\mathrm{iid}}{\sim}Q$ be mutually independent. The energy distance is defined as
\begin{equation}
\mathcal{E}(P,Q)
=
2E\|\mathbf{S}-\mathbf{T}\|_2
-
E\|\mathbf{S}-\mathbf{S}'\|_2
-
E\|\mathbf{T}-\mathbf{T}'\|_2 ,
\label{eq:energy_distance}
\end{equation}
and satisfies $\mathcal{E}(P,Q)\geq 0$ with equality if and only if $P=Q$ \citep{szekely2013energy}. Minimizing $\mathcal{E}$ therefore reduces discrepancies between entire distributions without designating individual moments to be balanced. In finite samples, however, minimizing the empirical energy distance should be understood as reducing an empirical distributional discrepancy, not as directly controlling bias for an arbitrary outcome regression.

At the sample level, consider estimating $E[Y(a)]$ by weighting individuals in group $A=a$. Let $F_{n,a,w}=\sum_{i\in\mathcal{I}_a}(w_i/n_a)\,\delta_{X_i}$ denote the weighted empirical distribution of $X$ in group $A=a$, where $\delta$ is the Dirac measure, and let $F_n$ be the (unweighted) empirical distribution of $X$ in the full sample. The empirical energy distance between them is
\begin{equation}
\begin{aligned}
\mathcal{E}(F_{n,a,w},F_n)
=\;&
\frac{2}{n_an}\sum_{i\in\mathcal{I}_a}\sum_{j=1}^{n}
w_i\|X_i-X_j\|_2\\
&-\frac{1}{n_a^2}\sum_{i\in\mathcal{I}_a}\sum_{j\in\mathcal{I}_a}
w_iw_j\|X_i-X_j\|_2\\
&-\frac{1}{n^2}\sum_{i=1}^{n}\sum_{j=1}^{n}
\|X_i-X_j\|_2 .
\end{aligned}
\label{eq:weighted_energy_ate}
\end{equation}
The third term does not involve the weights, so after dropping it, the objective is quadratic in $w$, and EBW is obtained as the solution to the quadratic programming problem
\begin{equation}
\begin{aligned}
\min_{w^{a,a}} \quad
&\mathcal{E}(F_{n,a,w^{a,a}},F_n)\\
\text{s.t.}\quad
&\sum_{i\in\mathcal{I}_a}w_i^{a,a}=n_a,\quad w_i^{a,a}\geq 0,\quad i\in\mathcal{I}_a .
\end{aligned}
\label{eq:ebw_ate}
\end{equation}
Note that the weights are normalized to sum to $n_a$, following the convention of the energy balancing literature, whereas the moment balancing literature commonly normalizes them to sum to one; the two conventions differ only by a scale factor and yield identical weighted means. Unlike \eqref{eq:mw_ate} and \eqref{eq:mw_mediation}, problem \eqref{eq:ebw_ate} involves neither basis functions nor tolerance parameters.

\cite{huling2024energy} also proposed an improved variant (iEBW), which augments the objective with the energy distance between the two weighted treatment-group distributions, in addition to the distance between each group and the full-sample target. This variant controls between-group balance more directly and has been reported to yield more stable finite-sample performance than standard EBW.

EBW was developed for balancing the covariate distribution in treatment effect estimation. In mediation analysis, however, the relevant target is not the marginal covariate distribution but the joint distribution of $V=(M,X)$ appearing in the mediation formula. Extending EBW therefore requires moving from covariate balancing for $\mu_{a,a}=E[Y(a)]$ to joint distribution balancing for counterfactual means such as $\mu_{1,0}=E[Y(1,M(0))]$, which is the subject of the next section.


\section{Energy Balancing Weights for Mediation Analysis}
\label{sec4}

\subsection{Basic Formulation of EBWMA}
We now formulate Energy Balancing Weights for Mediation Analysis (EBWMA). The goal is to construct, at the level of empirical distributions, the target joint distribution of $V=(M,X)$ required for counterfactual means such as
\[
\mu_{1,0}=\int g_1(m,x) \, dF_0^\star(m,x),
\qquad
dF_0^\star(m,x)=dF_{M\mid A=0,X=x}(m) \, dF_X(x).
\]
For clarity, we describe the construction for $\mu_{1,0}$. The same argument applies to other choices of $(a,a')$, with the corresponding treatment groups interchanged.

EBWMA proceeds in two stages. In the first stage, we construct $w^{0,0}$ and $w^{1,1}$ using the covariates $X$ alone, so that the weighted covariate distributions in the two treatment groups approximate the marginal distribution $F_X$. Under standard EBW, the weights for each group are obtained separately by solving \eqref{eq:ebw_ate} for $a=0$ and $a=1$. Under iEBW, the two sets of weights are obtained jointly by additionally penalizing the energy distance between the two weighted group distributions. We use iEBW throughout, since it has been reported to improve finite-sample performance over standard EBW \citep{huling2024energy}.

In the second stage, we construct the mediation weights $w^{1,0}$. Fixing the first-stage control weights $w^{0,0}$, EBWMA solves
\begin{equation}
\begin{aligned}
\min_{w^{1,0}} \quad
&\mathcal{E}\left(F_{n,1,w^{1,0}}^{V}, F_{n,0,w^{0,0}}^{V}\right) \\
\text{s.t.}\quad
&\sum_{i\in\mathcal{I}_1}w_i^{1,0}=n_1,\quad
w_i^{1,0}\geq 0,\qquad i\in\mathcal{I}_1,
\end{aligned}
\label{eq:ebwma_optim}
\end{equation}
where $F_{n,1,w^{1,0}}^{V}$ and $F_{n,0,w^{0,0}}^{V}$ denote the weighted empirical distributions of $V$ among individuals with $A=1$ and $A=0$, respectively. The second-stage weights are therefore chosen so that the weighted joint distribution of $(M,X)$ among treated individuals approximates the first-stage weighted joint distribution among controls. The first-stage weighted control distribution thus serves as the empirical target for the second-stage balancing problem and approximates $F_0^\star$.

As reflected in the constraints of \eqref{eq:ebwma_optim}, and following the convention of the energy balancing literature \citep{huling2024energy}, all EBWMA weights in this section are normalized to sum to the corresponding group size rather than to one, as in \eqref{eq:mw_ate} and \eqref{eq:mw_mediation}. The two normalization conventions differ only by a scale factor, and the self-normalized estimators in Section~\ref{sec4}.2 are invariant to this choice.

Expanding the objective in \eqref{eq:ebwma_optim} gives
\begin{equation}
\begin{aligned}
\mathcal{E}\left(F_{n,1,w^{1,0}}^{V},\,F_{n,0,w^{0,0}}^{V}\right)
=\;&
\frac{2}{n_1n_0}\sum_{i\in\mathcal{I}_1}\sum_{j\in\mathcal{I}_0}
w_i^{1,0}w_j^{0,0}\|V_i-V_j\|_2\\
&-\frac{1}{n_1^2}\sum_{i\in\mathcal{I}_1}\sum_{j\in\mathcal{I}_1}
w_i^{1,0}w_j^{1,0}\|V_i-V_j\|_2\\
&-\frac{1}{n_0^2}\sum_{i\in\mathcal{I}_0}\sum_{j\in\mathcal{I}_0}
w_i^{0,0}w_j^{0,0}\|V_i-V_j\|_2 .
\end{aligned}
\label{eq:ebwma_energy}
\end{equation}
The first two terms depend on $w^{1,0}$: the first is linear and rewards proximity between treated and (first-stage weighted) control observations, while the second is quadratic and discourages concentration of the weighted treated distribution, counteracting the tendency to place excessive weight on a few observations close to the target. The third term involves only the fixed first-stage weights and is therefore constant with respect to \eqref{eq:ebwma_optim}. Dropping this constant, the second stage reduces to a quadratic programming problem, which we solve with the osqp    solver \citep{stellato2018osqp,osqpPackage}. As in the original EBW, the formulation requires no hyperparameter tuning; the resulting weights may nonetheless depend on the solver configuration and on preprocessing choices, which we discuss in Section~\ref{sec4}.4.

This formulation replaces the finite-dimensional balance constraints
\[
\sum_{i\in\mathcal{I}_1} w_i^{1,0} v_k(M_i,X_i)
\approx
\sum_{i\in\mathcal{I}_0} w_i^{0,0} v_k(M_i,X_i),
\qquad k=1,\ldots,K,
\]
of moment balancing methods \citep{chan2016efficient,kawato2025balancing} with the minimization of an empirical distributional discrepancy between weighted distributions of $V$. EBWMA therefore requires no decision about which transformations or interactions of $(M,X)$ to specify as balance functions.

\subsection{Estimators via EBWMA}

Using $w^{0,0}$ and $w^{1,1}$ from the first stage and $w^{1,0}$ from \eqref{eq:ebwma_optim}, the counterfactual means are estimated by the self-normalized weighted averages
\begin{align}
\widehat{\mu}_{0,0}
&=\frac{\sum_{i\in\mathcal{I}_0}w_i^{0,0}Y_i}{\sum_{i\in\mathcal{I}_0}w_i^{0,0}},
\label{eq:ebwma_mu00}\\
\widehat{\mu}_{1,1}
&=\frac{\sum_{i\in\mathcal{I}_1}w_i^{1,1}Y_i}{\sum_{i\in\mathcal{I}_1}w_i^{1,1}},
\label{eq:ebwma_mu11}\\
\widehat{\mu}_{1,0}
&=\frac{\sum_{i\in\mathcal{I}_1}w_i^{1,0}Y_i}{\sum_{i\in\mathcal{I}_1}w_i^{1,0}},
\label{eq:ebwma_mu10}
\end{align}
which are invariant to the normalization convention adopted for the weights. The natural direct effect, natural indirect effect, and total effect are then estimated by
\begin{align}
\widehat{\tau}_{\mathrm{NDE}}
&=\widehat{\mu}_{1,0}-\widehat{\mu}_{0,0},
\label{eq:ebwma_nde}\\
\widehat{\tau}_{\mathrm{NIE}}
&=\widehat{\mu}_{1,1}-\widehat{\mu}_{1,0},
\label{eq:ebwma_nie}\\
\widehat{\tau}_{\mathrm{TE}}
&=\widehat{\mu}_{1,1}-\widehat{\mu}_{0,0},
\label{eq:ebwma_te}
\end{align}
so that the decomposition $\widehat{\tau}_{\mathrm{TE}}=\widehat{\tau}_{\mathrm{NDE}}+\widehat{\tau}_{\mathrm{NIE}}$ holds exactly in finite samples.

\subsection{Interpretation as Distribution Balancing}

The two-stage formulation can be understood as an empirical procedure for constructing the target joint distribution appearing in the mediation formula. We give the argument at the population level, where it is transparent; the procedure of Section~\ref{sec4}.1 is its empirical counterpart.

The first-stage control weights are intended to shift the covariate distribution among controls toward the marginal distribution $F_X$ of the target population. At the population level, they correspond to the density ratio
\[
w^{0,0}(x)\propto\frac{dF_X(x)}{dF_{X\mid A=0}(x)}.
\]
Because this ratio is a function of $X$ alone, reweighting by it alters the marginal distribution of $X$ while leaving the conditional law of $M$ given $(A=0,X)$ unchanged. When densities exist,
\begin{equation}
\begin{aligned}
w^{0,0}(x)\,dF_{M,X\mid A=0}(m,x)
&=
w^{0,0}(x)\,dF_{M\mid A=0,X=x}(m)\,dF_{X\mid A=0}(x)\\
&\propto
dF_{M\mid A=0,X=x}(m)\,dF_X(x)
\;=\;
dF_0^\star(m,x),
\end{aligned}
\label{eq:rn_derivation}
\end{equation}
so the weighted joint law of $V$ among controls is exactly the target $F_0^\star$. This is the formal statement behind the heuristic given in Section~\ref{sec3}.2, and it is what makes a first stage based on $X$ alone sufficient to define a target for the joint distribution of $V$. In finite samples, the first-stage weighted empirical distribution $F_{n,0,w^{0,0}}^{V}$ therefore serves as an empirical approximation to $F_0^\star$, subject to sampling variability and to the approximation error of the energy balancing step.

The second stage constructs weights $w^{1,0}$ among individuals with $A=1$ so that their weighted empirical distribution of $V$ approaches this empirical target. The resulting estimator $\widehat{\mu}_{1,0}$ thus draws outcomes from the treatment group, as required by $g_1(m,x)$, while reproducing the joint distribution of $(M,X)$ corresponding to $F_0^\star$. This is the same target distribution underlying ratio-weighting and $g$-formula estimators \citep{imai2010identification,vanderweele2009conceptual}; EBWMA differs in approximating it by minimizing an energy distance between empirical distributions, rather than by estimating the treatment assignment model or the conditional mediator density.

Two features of this argument should be noted. First, it holds at the population level, where the first-stage weights coincide with the density ratio above; the finite-sample weights only approximate this ratio, and the resulting error propagates into the second stage. Second, the second-stage objective measures discrepancy in the metric geometry of the space in which $V$ lives, and does not use the outcome regression $g_1$. The resulting bias in $\widehat{\mu}_{1,0}$ therefore depends on whether the residual distributional discrepancy is small in directions to which $g_1$ is sensitive.

\subsection{Implementation Practicalities and Algorithm}

Because the energy distance is built on the Euclidean metric, EBWMA depends on the scale and geometry of the variables collected in $V$. Preprocessing therefore determines which discrepancies the optimization treats as large, and is not a purely numerical detail. To prevent variables measured on large scales from dominating the pairwise distances, we standardize the mediator and covariates to zero mean and unit variance in all numerical analyses. Under severe skewness or heavy tails, further transformations may be preferable, since a few extreme observations can otherwise account for much of the empirical energy distance.

As a possible regularization strategy when the second-stage weights become highly concentrated or the optimization is numerically unstable, a ridge-type penalty
\[
\frac{\lambda}{2}\sum_{i\in\mathcal{I}_1}\left(w_i^{1,0}\right)^2
\]
can be added to the second-stage objective, as considered by \cite{huling2024energy}. This penalty discourages weight concentration and trades distributional balance against weight dispersion, at the cost of introducing the tuning parameter $\lambda$. We use the unpenalized formulation with $\lambda=0$ throughout the analyses reported here.

Weight diagnostics cannot establish that EBWMA is appropriate for a given dataset, but they can reveal substantial weight concentration and potential instability. We recommend inspecting, for the second-stage weights $w^{1,0}$,
\begin{itemize}
\item the \textbf{effective sample size},
\[
\mathrm{ESS}=
\frac{\left(\sum_{i\in\mathcal{I}_1} w_i^{1,0}\right)^2}
{\sum_{i\in\mathcal{I}_1} \left(w_i^{1,0}\right)^2},
\]
which equals $n_1$ when the weights are uniform and decreases as they concentrate;
\item the \textbf{relative maximum weight},
\[
\frac{\max_{i\in\mathcal{I}_1} w_i^{1,0}}
{\sum_{i\in\mathcal{I}_1} w_i^{1,0}},
\]
which quantifies the share of total weight assigned to the most heavily weighted observation;
\item the \textbf{coefficient of variation} of $\left(w_i^{1,0}\right)_{i\in\mathcal{I}_1}$, defined as its standard deviation divided by its mean.
\end{itemize}
All three are invariant to the normalization convention of Section~\ref{sec4}.1. A low ESS or a large maximum weight indicates that the estimate relies heavily on a limited subset of treated individuals and may therefore be more sensitive to perturbations of those observations.

The full procedure is summarized in Algorithm~\ref{alg:ebwma}.

\begin{algorithm}[htbp]
\caption{Energy Balancing Weights for Mediation Analysis (EBWMA)}
\label{alg:ebwma}
\begin{algorithmic}[1]
\Require Data $\{(X_i,A_i,M_i,Y_i)\}_{i=1}^{n}$
\Ensure $\widehat{\tau}_{\mathrm{NDE}}$, $\widehat{\tau}_{\mathrm{NIE}}$, $\widehat{\tau}_{\mathrm{TE}}$ and weight diagnostics
\State \textbf{Preprocessing.} Standardize $M$ and $X$, applying further transformations if warranted by the data structure; set $V_i=(M_i,X_i)$.
\State \textbf{First stage.} Obtain $w^{0,0}$ and $w^{1,1}$ from $X$: solve \eqref{eq:ebw_ate} separately for $a=0,1$ (EBW), or solve the corresponding joint problem for both groups (iEBW).
\State \textbf{Second stage.} Holding $w^{0,0}$ fixed, solve \eqref{eq:ebwma_optim} on $V$ to obtain $w^{1,0}$.
\State \textbf{Estimation.} Compute $\widehat{\mu}_{0,0}$, $\widehat{\mu}_{1,1}$, and $\widehat{\mu}_{1,0}$ from \eqref{eq:ebwma_mu00}--\eqref{eq:ebwma_mu10}, then $\widehat{\tau}_{\mathrm{NDE}}$, $\widehat{\tau}_{\mathrm{NIE}}$, and $\widehat{\tau}_{\mathrm{TE}}$ from \eqref{eq:ebwma_nde}--\eqref{eq:ebwma_te}.
\State \textbf{Diagnostics.} Evaluate the ESS, relative maximum weight, and coefficient of variation of $w^{1,0}$.
\end{algorithmic}
\end{algorithm}


\section{Numerical Simulation}
\label{sec5}

\subsection{Simulation Design}

This section evaluates whether EBWMA attains stable finite-sample performance in settings where the treatment and mediator models are difficult to specify parametrically. Four estimators were compared:
\begin{itemize}
\item glm: model-based weighting with logistic regression;
\item gbm: model-based weighting with gradient boosting, implemented in the twang package \citep{coffman2022tutorial};
\item 2-step MW: moment balancing via two-step minimal weights \citep{kawato2025balancing};
\item EBWMA-iEBW: the proposed method, with iEBW in the first stage.
\end{itemize}
Because we adopt iEBW throughout, we write EBWMA-iEBW for this implementation in the remainder of the paper.

For glm and gbm, the weights were formed from the second expression in \eqref{eq:ratio_weight_general}, with each component replaced by its fitted value; for $a=a'$ these reduce to standard inverse probability weights. For 2-step MW, we used $h(w)=w^2/2$. In DGP 1, first- and second-order terms entered the balance constraints with tolerance $\delta_k=0.05$. In DGP 2, the optimization frequently failed to converge under that configuration, so only first-order terms were included and the tolerance was relaxed to $\delta_k=0.10$. The hyperparameters for gbm followed \cite{coffman2022tutorial} without per-DGP tuning, since the performance of machine learning-based weighting estimators depends heavily on tuning and we sought a fixed implementation framework. EBWMA-iEBW required no tuning beyond the standardization described in Section~\ref{sec4}.4.

Two nonlinear data-generating processes (DGPs) were specified: DGP 1 follows the nonlinear transformation design of \cite{kang2007demystifying}, and DGP 2 the mixed continuous--binary covariate design of \cite{hainmueller2012entropy}. For each DGP we crossed the sample size $n\in\{500,1000\}$ with the treatment prevalence $\Pr(A=1)\in\{30\%,50\%,70\%\}$, giving twelve conditions, and generated 2{,}000 Monte Carlo replicates per condition. Performance was summarized by the bias, root mean squared error (RMSE), and Monte Carlo standard deviation (MCSD) of the TE, NDE, and NIE.

Even after relaxing the tolerance, 2-step MW failed to converge in a subset of replicates under DGP 2; all reported metrics for this method are therefore computed over converged replicates only, with convergence rates given in Table~\ref{tbl:sim_results_conv} in the appendix. This selective computation may flatter the apparent performance of 2-step MW, and comparisons involving it under DGP 2 should be read with that in mind. At the same time, the convergence difficulty is itself informative: it illustrates the practical cost of having to choose basis functions and tolerances under a complex nonlinear DGP, which is one of the motivations for EBWMA.

\subsection{Data-Generating Process 1}

DGP 1 extends the nonlinear transformation design of \cite{kang2007demystifying} to a mediation setting. In the original design, the observed covariates are nonlinear transformations of latent Gaussian variables, so that conventional main-effects regression models fitted to the observed covariates are misspecified. We added a post-treatment mediator and let the outcome depend on both the mediator and the latent covariate structure.

Latent variables $Z_1,\ldots,Z_4$ were drawn independently from $N(0,1)$, and the observed covariates were defined as
\begin{align*}
X_1 &= \exp(Z_1/2),\\
X_2 &= \frac{Z_2}{1+\exp(Z_1)},\\
X_3 &= \left(\frac{Z_1Z_3}{25}+0.6\right)^3,\\
X_4 &= (Z_2+Z_4+20)^2 .
\end{align*}
The treatment, mediator, and outcome mechanisms were specified through the latent variables as
\begin{align*}
\eta_A &= b_A - Z_1 + 0.5Z_2 - 0.25Z_3 - 0.1Z_4,\\
\eta_M &= 13.0Z_1 - 6.5Z_2 - 6.5Z_3 - 13.0Z_4,\\
\eta_Y &= 27.4Z_1 + 13.7Z_2 + 13.7Z_3 + 13.7Z_4 .
\end{align*}
With mutually independent standard normal errors $\varepsilon_M,\varepsilon_Y$,
the potential mediators and potential outcomes were generated as
\begin{align*}
M(a) &= 5a + \eta_M + \varepsilon_M, \\
Y(a,M(a')) &= 200 + 10a + M(a') + \eta_Y + \varepsilon_Y,
\end{align*}
and the observed variables as
\begin{align*}
A &\sim \mathrm{Bernoulli}\left\{\frac{\exp(\eta_A)}{1+\exp(\eta_A)}\right\},\qquad
M = AM(1)+(1-A)M(0),\\
Y &= AY(1,M(1))+(1-A)Y(0,M(0)).
\end{align*}
All methods were supplied with the observed covariates $(X_1,\ldots,X_4)$; the latent $Z_j$ were not available to the analyst. The intercept $b_A$ was set to $-1.1$, $0$, or $1.1$, yielding treatment prevalences of approximately $30\%$, $50\%$, and $70\%$. The true effects are $\mathrm{TE}=15$, $\mathrm{NDE}=10$, and $\mathrm{NIE}=5$.

\subsection{Data-Generating Process 2}

DGP 2 follows the covariate structure of \cite{hainmueller2012entropy} and combines correlated normal, uniform, skewed, and binary covariates:
\begin{align*}
(X_1,X_2,X_3)^\top
& \sim
N\left[
\begin{pmatrix}0\\0\\0\end{pmatrix},
\begin{pmatrix}
2 & 1 & -1\\
1 & 1 & -0.5\\
-1 & -0.5 & 1
\end{pmatrix}
\right],\\
X_4 &\sim \mathrm{Uniform}(-3,3),\qquad
X_5 \sim \chi^2(1),\qquad
X_6 \sim \mathrm{Bernoulli}(0.5).
\end{align*}
The heavily skewed $X_5$ and the binary $X_6$ make this a setting in which parametric specification and distance-based balancing are both sensitive to the geometry of the observed variables, though for different reasons.

The treatment, mediator, and outcome mechanisms were specified as
\begin{align*}
\eta_A &= b_A + X_1 + 2X_2 - 2X_3 - X_4 - 0.5X_5 + X_6,\\
\eta_M &= 0.2X_3X_4 + \sqrt{X_5},\qquad
\eta_Y = (X_1+X_2+X_5)^2 ,
\end{align*}
so that the mediator model contains an interaction and a square-root term and the outcome model a squared term. With mutually independent standard normal errors as above,
\begin{align*}
M(a) &= 5aX_6 + \eta_M + \varepsilon_M, \\
Y(a,M(a')) &= 5aX_5 + \eta_Y + M(a') + \varepsilon_Y
\end{align*}
and the observed variables were generated as
\begin{align*}
A &\sim \mathrm{Bernoulli}\{\Phi(\eta_A)\},\qquad
M = AM(1)+(1-A)M(0),\\
Y &= AY(1,M(1))+(1-A)Y(0,M(0)),
\end{align*}
where $\Phi$ is the standard normal distribution function. Setting $b_A$ to $-4$, $0$, or $4$ again gave treatment prevalences of approximately $30\%$, $50\%$, and $70\%$. The true effects are $\mathrm{TE}=7.5$, $\mathrm{NDE}=5.0$, and $\mathrm{NIE}=2.5$.

For the distance-based methods, $M$ and all covariates were standardized as described in Section~\ref{sec4}.4. The binary $X_6$ was standardized in the same way as the continuous variables; because standardization rescales a binary variable by $\{p(1-p)\}^{-1/2}$, the weight it carries in the Euclidean metric depends on its prevalence, and no separate scaling rule was applied.

\subsection{Simulation Results}

RMSE, bias, and MCSD are reported in Tables~\ref{tbl:sim_results_rmse}, \ref{tbl:supp_sim_results_bias}, and \ref{tbl:supp_sim_results_mcsd}; weight diagnostics for $w^{1,0}$, convergence rates for 2-step MW, and results under two alternative outcome models appear in Appendix~\ref{app:simulation_tables}.

\begin{landscape}
\begin{table}[htbp]
\centering
\caption{RMSE comparison of TE, NDE, and NIE across different DGPs and treatment probabilities.}
\label{tbl:sim_results_rmse}
\vspace{2mm}
\small
\begin{tabular}{ ccc ccc ccc ccc }
    \toprule
    & & & \multicolumn{3}{c}{$\Pr(A=1)=30\%$} & \multicolumn{3}{c}{$\Pr(A=1)=50\%$} & \multicolumn{3}{c}{$\Pr(A=1)=70\%$} \\
    \cmidrule(lr){4-6} \cmidrule(lr){7-9} \cmidrule(lr){10-12}
    DGP & $n$ & Method & TE & NDE & NIE & TE & NDE & NIE & TE & NDE & NIE \\
    \midrule
    \multirow{8}{*}{DGP 1} & \multirow{4}{*}{500}
      & glm        & 88.351 & 71.434 & 31.933 & 81.731 & 69.394 & 25.777 & 74.270 & 67.728 & 15.073 \\
    & & gbm        & 13.459 & 11.661 &  2.958 & 13.413 & 12.312 &  2.087 & 14.023 & 12.804 &  2.416 \\
    & & 2-step MW  &  8.625 &  4.161 &  5.260 & 10.157 &  4.765 &  5.674 & 10.817 &  5.533 &  5.570 \\
    & & EBWMA-iEBW &  2.996 &  1.225 &  3.113 &  2.791 &  2.063 &  4.541 &  3.431 &  2.787 &  5.916 \\
    \cmidrule(lr){2-12}
    & \multirow{4}{*}{1000}
      & glm        & 93.425 & 70.036 & 39.120 & 86.966 & 70.074 & 31.648 & 78.077 & 70.009 & 18.291 \\
    & & gbm        & 11.129 &  9.720 &  2.370 & 11.077 & 10.551 &  1.520 & 11.669 & 11.037 &  1.809 \\
    & & 2-step MW  &  8.726 &  4.055 &  5.067 & 10.341 &  4.937 &  5.578 & 11.034 &  5.668 &  5.522 \\
    & & EBWMA-iEBW &  2.056 &  2.024 &  3.879 &  1.928 &  3.272 &  5.099 &  2.378 &  3.903 &  6.166 \\
    \midrule
    \multirow{8}{*}{DGP 2} & \multirow{4}{*}{500}
      & glm        & 10.882 & 13.116 & 7.220 & 12.220 & 12.994 & 4.407 & 8.572 & 9.535 & 4.414 \\
    & & gbm        &  1.288 &  1.915 & 1.504 &  1.281 &  2.165 & 1.375 & 1.818 & 2.818 & 1.470 \\
    & & 2-step MW  &  2.246 &  3.332 & 3.763 &  1.504 &  3.028 & 3.463 & 1.670 & 3.633 & 2.966 \\
    & & EBWMA-iEBW &  0.717 &  1.063 & 0.714 &  0.568 &  0.737 & 0.525 & 0.788 & 0.718 & 0.532 \\
    \cmidrule(lr){2-12}
    & \multirow{4}{*}{1000}
      & glm        &  8.146 & 11.264 & 8.048 & 6.010 & 8.189 & 6.276 & 6.601 & 7.002 & 3.569 \\
    & & gbm        &  0.823 &  1.389 & 1.207 & 0.869 & 1.847 & 1.287 & 1.402 & 2.680 & 1.533 \\
    & & 2-step MW  &  2.218 &  2.677 & 3.465 & 1.331 & 2.663 & 3.412 & 1.166 & 3.331 & 2.792 \\
    & & EBWMA-iEBW &  0.473 &  0.785 & 0.583 & 0.376 & 0.529 & 0.425 & 0.501 & 0.485 & 0.423 \\
    \bottomrule
\end{tabular}
\vspace{1mm}
\begin{flushleft}
\footnotesize
DGP: Data Generating Process; 
TE: total effect; NDE: natural direct effect; NIE: natural indirect effect;
glm: generalized linear model; gbm: gradient boosting model; 2-step MW: 2-step minimal weights;
EBWMA-iEBW: energy balancing weights for mediation analysis (iEBW for first stage).
\end{flushleft}
\end{table}
\end{landscape}

\begin{landscape}
\begin{table}[htbp]
\centering
\caption{Bias comparison of TE, NDE, and NIE across different DGPs and treatment probabilities.}
\label{tbl:supp_sim_results_bias}
\vspace{2mm}
\small
\begin{tabular}{ ccc ccc ccc ccc }
    \toprule
    & & & \multicolumn{3}{c}{$\Pr(A=1)=30\%$} & \multicolumn{3}{c}{$\Pr(A=1)=50\%$} & \multicolumn{3}{c}{$\Pr(A=1)=70\%$} \\
    \cmidrule(lr){4-6} \cmidrule(lr){7-9} \cmidrule(lr){10-12}
    DGP & $n$ & Method & TE & NDE & NIE & TE & NDE & NIE & TE & NDE & NIE \\
    \midrule
    \multirow{8}{*}{DGP 1} & \multirow{4}{*}{500}
      & glm        & $-$79.973 & $-$54.971 & $-$25.002 & $-$73.226 & $-$54.023 & $-$19.202 & $-$66.321 & $-$56.564 & $-$9.758 \\
    & & gbm        & $-$13.101 & $-$11.136 &  $-$1.966 & $-$13.214 & $-$11.990 &  $-$1.223 & $-$13.709 & $-$12.282 & $-$1.426 \\
    & & 2-step MW  &  $-$7.850 &  $-$3.288 &  $-$4.562 &  $-$9.763 &  $-$4.446 &  $-$5.317 & $-$10.544 &  $-$5.258 & $-$5.286 \\
    & & EBWMA-iEBW &  $-$2.853 &     0.167 &  $-$3.020 &  $-$2.678 &     1.821 &  $-$4.499 &  $-$3.280 &     2.591 & $-$5.872 \\
    \cmidrule(lr){2-12}
    & \multirow{4}{*}{1000}
      & glm        & $-$85.796 & $-$53.005 & $-$32.791 & $-$79.325 & $-$54.547 & $-$24.778 & $-$70.266 & $-$57.928 & $-$12.339 \\
    & & gbm        & $-$10.949 &  $-$9.444 &  $-$1.504 & $-$10.964 & $-$10.361 &  $-$0.603 & $-$11.487 & $-$10.704 &  $-$0.783 \\
    & & 2-step MW  &  $-$8.255 &  $-$3.679 &  $-$4.576 & $-$10.150 &  $-$4.772 &  $-$5.378 & $-$10.905 &  $-$5.545 &  $-$5.360 \\
    & & EBWMA-iEBW &  $-$1.971 &     1.870 &  $-$3.842  & $-$1.865 &     3.218 &  $-$5.083 &  $-$2.295 &     3.853 &  $-$6.148 \\
    \midrule
    \multirow{8}{*}{DGP 2} & \multirow{4}{*}{500}
      & glm        & $-$5.715 & $-$5.480 & $-$0.235 & $-$5.010 & $-$4.778 & $-$0.233 & $-$4.835 & $-$4.245 & $-$0.590 \\
    & & gbm        &    0.157 &    0.563 & $-$0.406 &    0.625 &    1.457 & $-$0.832 &    1.073 &    2.145 & $-$1.072 \\
    & & 2-step MW  & $-$1.850 &    0.319 & $-$2.170 & $-$0.820 &    1.611 & $-$2.431 &    0.733 &    2.796 & $-$2.063 \\
    & & EBWMA-iEBW & $-$0.228 & $-$0.475 &    0.247 &    0.068 & $-$0.072 &    0.141 &    0.440 &    0.185 &    0.255 \\
    \cmidrule(lr){2-12}
    & \multirow{4}{*}{1000}
      & glm        & $-$5.353 & $-$4.396 & $-$0.957 & $-$4.645 & $-$3.599 & $-$1.046 & $-$4.784 & $-$3.937 & $-$0.846 \\
    & & gbm        & $-$0.047 &    0.481 & $-$0.528 &    0.498 &    1.445 & $-$0.947 &    1.050 &    2.297 & $-$1.248 \\
    & & 2-step MW  & $-$2.052 &    0.391 & $-$2.442 & $-$1.004 &    1.857 & $-$2.861 &    0.535 &    2.879 & $-$2.344 \\
    & & EBWMA-iEBW & $-$0.159 & $-$0.355 &    0.196 &    0.041 & $-$0.086 &    0.127 &    0.263 &    0.054 &    0.209 \\
    \bottomrule
\end{tabular}
\vspace{1mm}
\begin{flushleft}
\footnotesize
DGP: Data Generating Process; 
TE: total effect; NDE: natural direct effect; NIE: natural indirect effect;
glm: generalized linear model; gbm: gradient boosting model; 2-step MW: 2-step minimal weights;
EBWMA-iEBW: energy balancing weights for mediation analysis (iEBW for first stage).
\end{flushleft}
\end{table}

\begin{table}[htbp]
\centering
\caption{Monte Carlo standard deviation (MCSD) comparison of TE, NDE, and NIE across different DGPs and treatment probabilities.}
\label{tbl:supp_sim_results_mcsd}
\vspace{2mm}
\small
\begin{tabular}{ ccc ccc ccc ccc }
    \toprule
    & & & \multicolumn{3}{c}{$\Pr(A=1)=30\%$} & \multicolumn{3}{c}{$\Pr(A=1)=50\%$} & \multicolumn{3}{c}{$\Pr(A=1)=70\%$} \\
    \cmidrule(lr){4-6} \cmidrule(lr){7-9} \cmidrule(lr){10-12}
    DGP & $n$ & Method & TE & NDE & NIE & TE & NDE & NIE & TE & NDE & NIE \\
    \midrule
    \multirow{8}{*}{DGP 1} & \multirow{4}{*}{500}
      & glm        & 37.561 & 45.629 & 19.870 & 36.312 & 43.566 & 17.201 & 33.438 & 37.261 & 11.491 \\
    & & gbm        &  3.082 &  3.463 &  2.211 &  2.306 &  2.796 &  1.691 &  2.952 &  3.620 &  1.950 \\
    & & 2-step MW  &  3.575 &  2.551 &  2.620 &  2.802 &  1.715 &  1.983 &  2.413 &  1.724 &  1.757 \\
    & & EBWMA-iEBW &  0.916 &  1.213 &  0.754 &  0.784 &  0.970 &  0.615 &  1.004 &  1.027 &  0.720 \\
    \cmidrule(lr){2-12}
    & \multirow{4}{*}{1000}
      & glm        & 36.985 & 45.788 & 21.339 & 35.654 & 43.999 & 19.694 & 34.047 & 39.325 & 13.506 \\
    & & gbm        &  1.997 &  2.302 &  1.831 &  1.582 &  1.995 &  1.396 &  2.051 &  2.692 &  1.631 \\
    & & 2-step MW  &  2.828 &  1.707 &  2.176 &  1.982 &  1.267 &  1.481 &  1.684 &  1.172 &  1.328 \\
    & & EBWMA-iEBW &  0.584 &  0.773 &  0.534 &  0.490 &  0.594 &  0.414 &  0.623 &  0.622 &  0.466 \\
    \midrule
    \multirow{8}{*}{DGP 2} & \multirow{4}{*}{500}
      & glm        &  9.263 & 11.920 & 7.218 & 11.148 & 12.087 & 4.402 & 7.081 & 8.540 & 4.375 \\
    & & gbm        &  1.278 &  1.831 & 1.448 &  1.119 &  1.602 & 1.094 & 1.468 & 1.828 & 1.005 \\
    & & 2-step MW  &  1.274 &  3.318 & 3.076 &  1.260 &  2.565 & 2.466 & 1.502 & 2.320 & 2.132 \\
    & & EBWMA-iEBW &  0.680 &  0.951 & 0.670 &  0.564 &  0.733 & 0.506 & 0.654 & 0.694 & 0.467 \\
    \cmidrule(lr){2-12}
    & \multirow{4}{*}{1000}
      & glm        &  6.142 & 10.373 & 7.993 & 3.814 & 7.358 & 6.190 & 4.550 & 5.792 & 3.468 \\
    & & gbm        &  0.822 &  1.304 & 1.086 & 0.712 & 1.151 & 0.873 & 0.930 & 1.380 & 0.891 \\
    & & 2-step MW  &  0.842 &  2.649 & 2.459 & 0.873 & 1.909 & 1.859 & 1.036 & 1.675 & 1.517 \\
    & & EBWMA-iEBW &  0.445 &  0.700 & 0.549 & 0.374 & 0.522 & 0.406 & 0.427 & 0.482 & 0.368 \\
    \bottomrule
\end{tabular}
\vspace{1mm}
\begin{flushleft}
\footnotesize
DGP: Data Generating Process; 
TE: total effect; NDE: natural direct effect; NIE: natural indirect effect;
glm: generalized linear model; gbm: gradient boosting model; 2-step MW: 2-step minimal weights;
EBWMA-iEBW: energy balancing weights for mediation analysis (iEBW for first stage).
\end{flushleft}
\end{table}
\end{landscape}

Two main patterns emerge across the twelve conditions. First, glm performs far worse than every alternative, as expected: under DGP 1 the nonlinear transformations of \cite{kang2007demystifying} induce severe misspecification of the working logistic models, and its RMSE for the NDE exceeds 67 at every prevalence and sample size. Second, EBWMA-iEBW attains the smallest MCSD for all three estimands in all twelve simulation conditions, typically by a factor of two to four relative to gbm and 2-step MW. Whatever the bias behavior of the method, its sampling variability was uniformly the lowest among those compared.

Under DGP 2, EBWMA-iEBW attained the smallest RMSE for all three estimands in every condition, while maintaining absolute bias below 0.5 throughout; absolute bias ranged from 0.05 to 2.30 for gbm and from 0.32 to 2.88 for 2-step MW. For EBWMA-iEBW, increasing $n$ from 500 to 1{,}000 reduced MCSD in every case and absolute bias in almost every case, indicating improved finite-sample performance at the larger sample size. The 2-step MW figures for this DGP should be read with caution, however, since they are computed over converged replicates only and convergence was far from universal at $n=500$ (Table~\ref{tbl:sim_results_conv}).

Under DGP 1 the picture is more mixed and depends on the estimand. For the TE, EBWMA-iEBW again gave the smallest RMSE in all six conditions. For the NDE, it had the smallest absolute bias among the four methods in all six conditions, ranging from 0.167 to 3.853 against 9.444 to 12.282 for gbm, although its bias grew with both $n$ and treatment prevalence. For the NIE, by contrast, gbm outperformed EBWMA-iEBW in every condition on both RMSE and absolute bias; moreover, the bias of EBWMA-iEBW increased with $n$ at each prevalence ($-3.020$ to $-3.842$ at $\Pr(A=1)=30\%$, $-4.499$ to $-5.083$ at $50\%$, and $-5.872$ to $-6.148$ at $70\%$), so that its RMSE for the NIE was driven almost entirely by bias rather than by variability.

This pattern is the clearest limitation exposed by the simulations, and it is consistent with the discussion in Section~\ref{sec4}.3. The second-stage objective measures discrepancy in the Euclidean geometry of the standardized $(M,X)$ space and does not use the outcome regression. Under DGP 1 the observed covariates are strongly nonlinear transformations of the latent variables: $X_1=\exp(Z_1/2)$ is lognormal and $X_4=(Z_2+Z_4+20)^2$ is highly asymmetric, and standardization removes neither the skewness nor the tail behavior. The energy distance can then be reduced while leaving residual discrepancies in directions that matter for $g_1$. Both the NDE and the NIE involve $\widehat\mu_{1,0}$ and are therefore exposed to any residual imbalance in the second-stage reconstruction of $F_0^\star$; the TE, estimated by $\widehat\mu_{1,1}-\widehat\mu_{0,0}$, does not involve $w^{1,0}$ at all, which is consistent with its uniformly small RMSE here. That the bias did not decrease between $n=500$ and $n=1{,}000$ suggests it is not attributable to sampling variability alone, and may instead reflect a persistent difficulty in reconstructing outcome-relevant features of the target distribution under this covariate geometry.

The alternative outcome models reported in Appendix~\ref{app:simulation_tables} bear directly on this reading, since they hold the treatment and mediator mechanisms---and hence the target distribution $F_0^\star$---fixed and vary only the regression through which residual imbalance is integrated. Under DGP 1, the ratio of the NIE RMSE of EBWMA-iEBW to that of gbm ranged from 1.05 to 3.41 under the primary outcome model and from 1.28 to 2.86  under a threshold outcome model, but from 0.95 to 1.33 under an outcome model multiplicative in the mediator, where EBWMA-iEBW was the more accurate of the two at $n=1{,}000$ with $\Pr(A=1)=50\%$ and $70\%$. The disadvantage for the NIE is therefore not a fixed property of this covariate geometry: it varies with the outcome regression while the estimation target is held constant, which is what a criterion indifferent to $g_a$ would predict. The same comparison also exposes a further sensitivity, in that 2-step MW attained the smallest NDE RMSE in every condition under the threshold model. Under DGP 2, by contrast, EBWMA-iEBW retained the smallest RMSE for all three estimands under both alternatives, with a single marginal exception, so its advantage there is not an artifact of the outcome specification. These observations motivate the preprocessing and regularization strategies noted in Section~\ref{sec4}.4 and, more fundamentally, the incorporation of outcome information discussed in Section~\ref{sec7}.

Treatment prevalence affected total and mediation effect estimations through distinct mechanisms. Across all data-generating processes, outcome specifications, and sample sizes, total effect (TE) estimation consistently attained its lowest RMSE at $\Pr(A=1) = 50\%$. For the NDE and NIE, the impact of prevalence depended on covariate distribution and outcome geometry. Under DGP 2, where covariate distributions are relatively unskewed, balanced or higher prevalence ($\Pr(A=1) \ge 50\%$) consistently yielded smaller RMSEs. Conversely, under DGP 1, where covariate distributions are skewed, the propagation of residual imbalance depended heavily on the functional form of the outcome surface. Consequently, lower prevalence ($\Pr(A=1) = 30\%$) yielded the lowest RMSE under the primary outcome specification (Model 1), whereas higher prevalence ($\Pr(A=1) = 70\%$) was optimal for NIE under the multiplicative outcome specification (Model 2; see Appendix~\ref{app:altoutcome} for details).

The weight diagnostics qualify these results. EBWMA-iEBW had a lower effective sample size, a larger relative maximum weight, and a higher coefficient of variation than gbm in every condition; under DGP 1 with $\Pr(A=1)=70\%$ and $n=1{,}000$, its median effective sample size was 234 against 649 for gbm, or roughly a third of $n_1=700$. Its low Monte Carlo variability therefore did not arise from more uniform weights. These diagnostics do not admit the usual propensity-score interpretation here, since the weights solve a distributional balancing problem rather than inverting an estimated assignment probability. They nevertheless characterize the degree of weight concentration: a low effective sample size or a large maximum weight indicates that the estimator relies heavily on a limited subset of treated individuals and may therefore be more sensitive to perturbations of those observations. The favorable RMSE figures should be interpreted alongside these diagnostics.

Overall, the simulations indicate that EBWMA is a useful alternative when parametric treatment and mediator models are difficult to specify, and that its sampling variability was uniformly the lowest among the methods compared, albeit with more concentrated weights. Its bias behavior depended jointly on the covariate geometry and the outcome regression: favorable under the mixed continuous--binary structure of DGP 2 across all three outcome specifications considered, but, for the NIE under the nonlinear transformations of DGP 1, unfavorable under two of the three and sensitive to which was used.


\section{Real Data Analysis}
\label{sec6}

\subsection{Data Description}

We illustrate EBWMA using the National Health and Nutrition Examination Survey I Epidemiologic Follow-up Study (NHEFS), a common benchmark in causal inference research \citep{hernan2020causal,inoue2024confounder}. Following the data structure of \cite{inoue2024confounder}, we used baseline variables measured in 1971, body weight measured in 1982, and all-cause mortality through 1992. A complete-case analysis left $n=1{,}507$ individuals.

The treatment $A$ was a binary indicator of baseline annual household income in 1971, with $A=1$ for incomes of \$20{,}000 or more and $A=0$ otherwise. The mediator $M$ was body weight in kilograms in 1982, and the outcome $Y$ was all-cause mortality through 1992. Baseline age, sex, race, and physical activity level served as adjustment variables, with discrete covariates dummy-coded; the mediator and all covariates were standardized as in Section~\ref{sec4}.4. The estimands were the TE, NDE, and NIE of baseline income on mortality, on the risk difference scale.

EBWMA-iEBW was compared with the same three estimators used in the simulations. For glm and gbm, the weights were formed from the second expression in \eqref{eq:ratio_weight_general}, with each component replaced by its fitted value. For 2-step MW, we used $h(w)=w^2/2$ with first- and second-order terms for the continuous variables in the balance constraints. Because no principled choice of tolerance is available, we report results for both $\delta_k=0.05$ and $\delta_k=0.10$, as in the simulations.

This analysis is an illustrative application rather than a substantive investigation of income, body weight, and mortality. Conditions (B1)--(B4) are untestable and implausible in several respects here: income is dichotomized at a single threshold, the covariate set is deliberately small, and the analysis is restricted to individuals who survived to the 1982 follow-up and had body weight recorded, which is itself a post-treatment selection. The estimates below should be read accordingly.

\subsection{Analysis Results}

Table~\ref{tbl:real_data_table} reports the point estimates as risk differences in percentage points. All methods indicate lower mortality in the higher-income group, but the magnitude of the TE varies more than threefold across methods: $-12.86$ for glm against $-5.85$ for EBWMA-iEBW, with gbm ($-6.11$) close to the latter and 2-step MW intermediate. This spread mirrors the simulations, where glm was the most biased method under both DGPs, though with a single dataset we cannot attribute the discrepancy to misspecification.

The estimated NIE was small relative to the estimated TE for every method, so the fitted decomposition assigned little of the total contrast to the body-weight pathway. Its sign, however, was not consistent: glm and gbm gave small positive values ($0.05$ and $0.06$), whereas the balancing methods gave negative ones ($-0.55$ to $-0.99$). The 2-step MW estimates also shifted appreciably between the two tolerances, changing the TE by 1.12 percentage points, which illustrates the sensitivity to a choice that the method leaves to the analyst. The NIE estimates are small in absolute terms, but their variation across methods shows that the mediation component is the part of the decomposition most sensitive to the weighting approach—consistent with the DGP 1 results in Section~\ref{sec5}.4.

We report point estimates only. Interval estimation for EBWMA requires a bootstrap whose validity for this two-stage optimization-based estimator has not been established, as discussed in Section~\ref{sec7}; the comparisons above should therefore be read as descriptive.

\begin{table}[htbp]
	\centering
	\caption{Estimated effects in the illustrative NHEFS mediation analysis.}
	\label{tbl:real_data_table}
    \vspace{2mm}
    \begin{tabular*}{\linewidth}{@{\extracolsep{\fill}} c ccc @{}}
		\toprule
		& \multicolumn{3}{c}{Point Estimate [\%]}\\
		\cmidrule(lr){2-4}
		Method & TE & NDE & NIE \\
		\midrule
		glm                         & $-$12.86 & $-$12.91 &    0.05 \\
		gbm                         &  $-$6.11 &  $-$6.17 &    0.06 \\
		2-step MW ($\delta_k=0.05$) &  $-$7.60 &  $-$6.68 & $-$0.92 \\
		2-step MW ($\delta_k=0.10$) &  $-$8.72 &  $-$7.73 & $-$0.99 \\
		EBWMA-iEBW                  &  $-$5.85 &  $-$5.31 & $-$0.55 \\
		\bottomrule
	\end{tabular*}

	\vspace{1mm}
	\begin{flushleft}
	\footnotesize
	Note: Estimates are risk differences expressed in percentage points. Because of rounding, TE may not exactly equal the sum of NDE and NIE. TE: total effect; NDE: natural direct effect; NIE: natural indirect effect; 2-step MW: 2-step minimal weights; EBWMA-iEBW: energy balancing weights for mediation analysis (iEBW for first stage).
	\end{flushleft}
\end{table}


\section{Discussion}
\label{sec7}

We proposed Energy Balancing Weights for Mediation Analysis (EBWMA), which estimates counterfactual means in causal mediation analysis by approximating the target joint distribution of the mediator and covariates. The representation
\[
\mu_{a,a'}=\int g_a(m,x) \, dF_{a'}^\star(m,x)
\]
makes explicit that weighting for natural effects requires reconstruction of a joint distribution that is generally not observed in either treatment arm. For example, estimating $\mu_{1,0}=E[Y(1,M(0))]$ requires outcomes from individuals with $A=1$ while reproducing the joint distribution of $(M,X)$ that combines the mediator distribution under $A=0$ with the marginal covariate distribution of the target population. From this perspective, density-ratio weighting, moment balancing, and EBWMA can be viewed as alternative ways of approximating the same target distribution: through estimated conditional probabilities or densities, finitely many prespecified balance functions, or a global empirical distributional discrepancy, respectively.

The two-stage construction of EBWMA follows naturally from this representation. In the first stage, weighting within the control group shifts the marginal covariate distribution toward $F_X$ while leaving the conditional mediator distribution given $(A=0,X)$ unchanged, thereby constructing an empirical approximation to $F_0^\star$. In the second stage, the treated group is reweighted so that its joint distribution of $(M,X)$ approaches this constructed target, allowing outcomes observed under $A=1$ to be averaged over the distribution required for $\mu_{1,0}$. The two stages therefore reflect distinct components of the mediation formula rather than merely a computational strategy. Unlike moment balancing, EBWMA requires neither prespecified basis functions nor tolerance parameters. This structure is also a source of uncertainty because the second-stage target is itself estimated in the first stage.

The simulations showed that the performance of EBWMA depends on the data-generating structure. EBWMA attained the smallest Monte Carlo variability in every condition considered. Under the mixed continuous--binary structure of DGP 2, it also gave the smallest RMSE for all three estimands in nearly every condition across the three outcome specifications. Under the nonlinear transformation structure of DGP 1, however, its performance was less uniform. It performed favorably for the TE and NDE, whereas gradient-boosted IPW gave smaller RMSE and absolute bias for the NIE under the primary outcome model, and the bias of EBWMA did not diminish between $n=500$ and $n=1{,}000$. When only the outcome regression was changed while the treatment and mediator mechanisms, and hence $F_0^\star$, were held fixed, this disadvantage narrowed or reversed. The NHEFS analysis further showed that EBWMA can be implemented in observational data and produced estimates broadly comparable with alternative weighting approaches, although that analysis was intended as an illustration rather than a substantive causal investigation.

These findings highlight a distinction between global distributional balance and balance relevant to the target estimand. EBWMA minimizes an overall discrepancy between weighted distributions of $(M,X)$, whereas the counterfactual mean depends on the target distribution through its integration against $g_a(m,x)$. Consequently, a small energy distance does not necessarily imply small bias if residual imbalance remains in features of $(M,X)$ that are important for the outcome regression. The outcome-model comparison in Appendix~\ref{app:simulation_tables} illustrates this directly: the relative performance of EBWMA changed when the outcome regression changed, even though the target distribution remained fixed. This issue is not specific to energy balancing. Moment balancing can face the same problem when the selected balance functions or tolerances fail to control features that are important for the outcome regression.

Several limitations should be noted. EBWMA does not relax the identification conditions for natural direct and indirect effects and therefore remains vulnerable to unmeasured confounding, treatment-induced mediator--outcome confounding, and other violations of the mediation formula. The energy-distance criterion also introduces its own substantive choices, since standardization and transformation determine the geometry of the mediator--covariate space and hence which discrepancies are emphasized; moment balancing replaces this dependence on a metric with an explicit choice of balance functions, and neither choice is innocuous. High dimensionality may further limit the informativeness of pairwise distances, and limited overlap can concentrate the weights. In our simulations, however, EBWMA combined the lowest Monte Carlo variability with lower effective sample sizes and larger maximum weights than gradient-boosted IPW, indicating that weight-concentration diagnostics remain important but should not be read as direct surrogates for sampling variance.

The numerical evidence is also limited in scope. Our comparisons involved only weighting-based estimators, two data-generating processes, moderate sample sizes, a single continuous mediator, and low-dimensional covariates, with no benchmarking against outcome-regression, doubly robust, or semiparametrically efficient mediation estimators \citep{tchetgen2012semiparametric,zheng2012targeted}. The theoretical properties of the two-stage estimator---consistency, convergence rates, asymptotic distributions, and formal inference---also remain open. The nonparametric bootstrap is a natural candidate, but its validity is unproven, alternatives developed for energy-balancing procedures may merit consideration \citep{santra2026distributional}, and repeating both optimization stages in each resample is computationally demanding. These limitations motivate regularization, scalable approximations to the energy distance, comparisons with efficient estimators, and outcome-informed balancing criteria as directions for further work.

In conclusion, EBWMA reformulates weighting for causal mediation analysis as empirical reconstruction of the target joint distribution required by the mediation formula. It complements rather than supersedes existing weighting and semiparametric approaches, replacing explicit mediator density modeling and prespecified moment conditions with a distribution-level balancing criterion. Its main attraction is this model-agnostic reconstruction of the target distribution; its main limitation is that global geometric balance need not coincide with balance in directions most relevant to the outcome regression.


\backmatter

\bmhead{Acknowledgements}

\section*{Declarations}
\begin{itemize}
\item \textbf{Funding:} TO received tuition support from Biostatistics, Asia Research and Development, GlaxoSmithKline. KS received no funding for this study.

\item \textbf{Competing interests:} TO is employed by GlaxoSmithKline K.K. and received tuition support from the company. This study was conducted as part of TO's doctoral research and was not designed for application to any specific clinical trial or product development program of the company. The company had no role in the study design, data collection, analysis, interpretation of results, or writing of the manuscript. KS declares no competing interests. The authors declare no competing interests related to this work.

\item \textbf{Ethics approval:} Not applicable.

\item \textbf{Consent to participate:} Not applicable.

\item \textbf{Data availability:} The NHEFS data used in Section~\ref{sec6} are available from the causaldata package in R.

\item \textbf{Code availability:} The code used in this study is available from the corresponding author upon reasonable request.

\item \textbf{Author contributions:} TO contributed to the study conception and design, performed the data analysis and computer programming, and wrote the first draft of the manuscript. KS supervised the research and reviewed and edited the manuscript. Both authors read and approved the final manuscript.
\end{itemize}

\bibliography{sn-bibliography}

\clearpage
\appendix
\section{Supplementary Tables for the Numerical Simulation and Real-Data Analysis}
\label{app:simulation_tables}

\subsection{Weight Diagnostics}
\label{app:diagnostics}

Tables~\ref{tbl:supp_sim_results_ess}--\ref{tbl:supp_sim_results_cv} summarize, across the 2{,}000 replicates of each condition, the effective sample size, relative maximum weight, and coefficient of variation of the second-stage weights $w^{1,0}$, comparing gbm with EBWMA-iEBW.

EBWMA-iEBW was less favorable than gbm on all three diagnostics in every condition. The gap widened with treatment prevalence: under DGP 1 with $\Pr(A=1)=70\%$ and $n=1{,}000$, the median effective sample size was 234 for EBWMA-iEBW against 649 for gbm, or roughly a third of $n_1=700$. Under DGP 2, where the treated and control mediator distributions differ by $5X_6$, both methods produced concentrated weights, but EBWMA-iEBW more so. As reported in Section~\ref{sec5}.4, this greater weight concentration nevertheless coexisted with uniformly lower Monte Carlo variability. The diagnostics should therefore be interpreted as measures of how strongly the estimator relies on a limited subset of observations, rather than as direct measures of its sampling variability.

\subsection{Convergence of the Two-Step Minimal Weights Method}
\label{app:convergence}

Table~\ref{tbl:sim_results_conv} reports the proportion of replicates in which the optimization for 2-step MW converged for all three weight vectors. Convergence was complete under DGP 1. Under DGP 2 it failed in a substantial fraction of replicates at the smaller sample size: only 47.4\% converged at $\Pr(A=1)=30\%$ with $n=500$, and 95.0\% at $50\%$. Because the reported metrics for 2-step MW are computed over converged replicates only, the DGP 2 figures for this method---particularly at $\Pr(A=1)=30\%$ with $n=500$---reflect a selected subset of replicates and are not directly comparable with the others.

\subsection{Alternative Outcome Models}
\label{app:altoutcome}

To assess whether the findings of Section~\ref{sec5}.4 are specific to the outcome models used there, we repeated the simulations under two alternative outcome specifications, holding the treatment and mediator mechanisms---and hence the target distribution $F_{a'}^\star$---fixed. Writing $(a,a')\in\{(1,1),(1,0),(0,0)\}$, the specifications for DGP 1 are

\begin{align*}
\text{model 2:}\quad
Y(a,M(a')) &= c_a + 3\,M(a')\,\eta_Y + \varepsilon_Y,\\
\text{model 3:}\quad
Y(a,M(a')) &= c_a + I\{M(a')>2.5\} + I\{\eta_Y>0\} + \varepsilon_Y,
\end{align*}
with $c_1=210$ and $c_0=200$, and those for DGP 2 are

\begin{align*}
\text{model 2:}\quad
Y(a,M(a')) &= 5aX_5 + \eta_Y \, M(a') + \varepsilon_Y,\\
\text{model 3:}\quad
Y(a,M(a')) &= 5aX_5 + \eta_Y + I\{M(a')>2.5\} + I\{\eta_Y>0\} + \varepsilon_Y.
\end{align*}
Model 2 makes the outcome multiplicative in the mediator and the latent index; model 3 replaces continuous dependence on the mediator and on the index with threshold indicators. Because the scale of $Y$ differs markedly across the three outcome specifications, RMSE values are comparable within a table but not across Tables~\ref{tbl:sim_results_rmse}, \ref{tbl:rmse_Y2_dgp1}--\ref{tbl:rmse_Y2_dgp2}, and \ref{tbl:rmse_Y3_dgp1}--\ref{tbl:rmse_Y3_dgp2}; the true effects are given in the table notes.

Results appear in Tables~\ref{tbl:rmse_Y2_dgp1}--\ref{tbl:rmse_Y3_dgp2}. Under DGP 2, EBWMA-iEBW attained the smallest RMSE for all three estimands in every condition under both alternatives, with the single exception of the NDE at $\Pr(A=1)=30\%$ with $n=1{,}000$ under model 2, where gbm was marginally better (4.445 against 4.477). The advantage reported in Section~\ref{sec5}.4 is therefore not an artifact of the outcome specification.

Under DGP 1 the outcome specification mattered, but only for model 2. Taking the ratio of the NIE RMSE of EBWMA-iEBW to that of gbm across the six conditions, this ratio ranged from 1.05 to 3.41 under the primary outcome model and from 1.28 to 2.86 under model 3, but from 0.95 to 1.33 under model 2, where EBWMA-iEBW attained the smaller RMSE at $n=1{,}000$ with $\Pr(A=1)=50\%$ (91.494 against 94.012) and $70\%$ (67.095 against 70.262). The disadvantage for the NIE thus narrowed substantially, and was reversed in two conditions, under an outcome model that is multiplicative in the mediator, while persisting at a similar magnitude under the threshold model. For the TE, EBWMA-iEBW retained the smallest RMSE in all six conditions under both alternatives; for the NDE it did so under model 2, whereas under model 3 2-step MW was smallest in every condition.

That the relative performance of EBWMA-iEBW changed while its estimation target did not is consistent with the interpretation offered in Section~\ref{sec7}: the second-stage criterion balances $(M,X)$ without reference to the outcome regression, so the consequences of a given residual imbalance depend on the regression surface through which that imbalance is integrated.

\subsection{Covariate Balance in tunskewedhe Real-Data Analysis}
\label{app:smd}

Table~\ref{tbl:supp_realdata_smd} reports standardized mean differences for the covariates and the mediator in the NHEFS analysis, computed against the pooled pre-weighting standard deviation. EBWMA-iEBW achieved the smallest absolute difference for most variables in both comparisons, the exceptions being race in the $w^{1,0}$ versus $w^{0,0}$ comparison, where 2-step MW was smaller. For the mediator, the $w^{1,0}$ versus $w^{0,0}$ difference was $0.00009$ for EBWMA-iEBW, against $0.01906$ for glm and $0.03988$ for gbm. The glm weights left age and race less balanced than they were before weighting ($0.289$ against $0.241$ unadjusted, and $0.473$ against $0.282$), which is consistent with the comparatively large glm estimates in Table~\ref{tbl:real_data_table}. Standardized mean differences compare first moments only, and therefore neither capture the distributional balance that energy balancing targets nor rule out residual imbalance in higher moments.


\begin{table}[htbp]
\centering
\caption{Effective sample size of $w^{1,0}$ across data-generating processes and treatment prevalences.}
\label{tbl:supp_sim_results_ess}
\vspace{2mm}
\small
\begin{tabular}{ cccccccccc }
    \toprule
    DGP & $\Pr(A=1)$ & $n$ & Method & Min & 1st Qu & Median & Mean & 3rd Qu & Max \\
    \midrule
    \multirow{12}{*}{DGP 1} & \multirow{4}{*}{$30\%$} & \multirow{2}{*}{500}
        & gbm        &  62.110 &  98.335 & 107.857 & 107.495 & 116.729 & 152.315 \\
    & & & EBWMA-iEBW &  42.504 &  62.533 &  68.439 &  68.301 &  73.831 &  98.029 \\
    \cmidrule(lr){3-10}
    & & \multirow{2}{*}{1000}
        & gbm        & 130.341 & 194.093 & 209.101 & 208.611 & 223.528 & 268.707 \\
    & & & EBWMA-iEBW &  74.745 & 109.599 & 116.422 & 116.644 & 123.918 & 148.139 \\
    \cmidrule(lr){2-10}
    & \multirow{4}{*}{$50\%$} & \multirow{2}{*}{500}
        & gbm        & 166.126 & 205.792 & 216.509 & 215.924 & 225.928 & 262.552 \\
    & & & EBWMA-iEBW &  81.497 & 106.391 & 112.231 & 112.134 & 118.140 & 145.893 \\
    \cmidrule(lr){3-10}
    & & \multirow{2}{*}{1000}
        & gbm        & 350.812 & 411.496 & 426.438 & 425.573 & 440.804 & 491.289 \\
    & & & EBWMA-iEBW & 156.832 & 187.734 & 195.785 & 195.770 & 203.568 & 238.820 \\
    \cmidrule(lr){2-10}
    & \multirow{4}{*}{$70\%$} & \multirow{2}{*}{500}
        & gbm        & 186.996 & 315.784 & 325.084 & 324.624 & 333.941 & 367.314 \\
    & & & EBWMA-iEBW &  82.251 & 121.960 & 130.047 & 129.924 & 137.664 & 166.495 \\
    \cmidrule(lr){3-10}
    & & \multirow{2}{*}{1000}
        & gbm        & 543.317 & 633.556 & 648.786 & 648.491 & 663.441 & 719.874 \\
    & & & EBWMA-iEBW & 176.361 & 221.773 & 233.585 & 233.128 & 244.443 & 290.527 \\
    \midrule
    \multirow{12}{*}{DGP 2} & \multirow{4}{*}{$30\%$} & \multirow{2}{*}{500}
        & gbm        &  15.168 &  43.774 &  51.008 &  50.978 &  58.208 &  92.338 \\
    & & & EBWMA-iEBW &  16.545 &  24.166 &  26.561 &  26.652 &  28.885 &  38.774 \\
    \cmidrule(lr){3-10}
    & & \multirow{2}{*}{1000}
        & gbm        &  32.309 &  81.735 &  92.414 &  92.579 & 102.914 & 149.782 \\
    & & & EBWMA-iEBW &  22.962 &  33.553 &  36.260 &  36.433 &  39.208 &  51.324 \\
    \cmidrule(lr){2-10}
    & \multirow{4}{*}{$50\%$} & \multirow{2}{*}{500}
        & gbm        &  46.700 & 102.861 & 113.981 & 113.287 & 124.339 & 184.493 \\
    & & & EBWMA-iEBW &  25.782 &  37.581 &  40.702 &  40.890 &  44.036 &  58.065 \\
    \cmidrule(lr){3-10}
    & & \multirow{2}{*}{1000}
        & gbm        & 100.405 & 184.004 & 203.698 & 201.866 & 221.076 & 291.003 \\
    & & & EBWMA-iEBW &  35.833 &  50.438 &  54.390 &  54.455 &  58.472 &  74.970 \\
    \cmidrule(lr){2-10}
    & \multirow{4}{*}{$70\%$} & \multirow{2}{*}{500}
        & gbm        &  75.258 & 173.224 & 190.673 & 188.570 & 205.942 & 271.052 \\
    & & & EBWMA-iEBW &  31.003 &  45.438 &  49.622 &  49.888 &  53.832 &  75.796 \\
    \cmidrule(lr){3-10}
    & & \multirow{2}{*}{1000}
        & gbm        &  63.576 & 306.712 & 342.233 & 336.212 & 371.032 & 488.652 \\
    & & & EBWMA-iEBW &  40.904 &  59.473 &  64.813 &  64.906 &  69.901 &  93.458 \\
    \bottomrule
\end{tabular}
\vspace{1mm}
\begin{flushleft}
\footnotesize
Note: Summary statistics are taken over the 2{,}000 replicates of each condition. For reference, the number of treated individuals is $n_1 = n \times \Pr(A=1)$ in expectation.
DGP: data-generating process; gbm: gradient boosting model;
EBWMA-iEBW: energy balancing weights for mediation analysis (iEBW for first stage);
1st Qu: first quartile; 3rd Qu: third quartile.
\end{flushleft}
\end{table}

\begin{table}[htbp]
\centering
\caption{Relative maximum weight of $w^{1,0}$ across data-generating processes and treatment prevalences.}
\label{tbl:supp_sim_results_rmw}
\vspace{2mm}
\small
\begin{tabular}{ cccccccccc }
    \toprule
    DGP & $\Pr(A=1)$ & $n$ & Method & Min & 1st Qu & Median & Mean & 3rd Qu & Max \\
    \midrule
    \multirow{12}{*}{DGP 1} & \multirow{4}{*}{$30\%$} & \multirow{2}{*}{500}
        & gbm        & 0.012 & 0.021 & 0.024 & 0.025 & 0.029 & 0.083 \\
    & & & EBWMA-iEBW & 0.019 & 0.031 & 0.036 & 0.037 & 0.041 & 0.076 \\
    \cmidrule(lr){3-10}
    & & \multirow{2}{*}{1000}
        & gbm        & 0.007 & 0.012 & 0.014 & 0.014 & 0.016 & 0.039 \\
    & & & EBWMA-iEBW & 0.014 & 0.021 & 0.023 & 0.024 & 0.026 & 0.045 \\
    \cmidrule(lr){2-10}
    & \multirow{4}{*}{$50\%$} & \multirow{2}{*}{500}
        & gbm        & 0.007 & 0.010 & 0.012 & 0.012 & 0.014 & 0.027 \\
    & & & EBWMA-iEBW & 0.014 & 0.020 & 0.022 & 0.022 & 0.024 & 0.046 \\
    \cmidrule(lr){3-10}
    & & \multirow{2}{*}{1000}
        & gbm        & 0.004 & 0.006 & 0.006 & 0.007 & 0.007 & 0.015 \\
    & & & EBWMA-iEBW & 0.010 & 0.013 & 0.014 & 0.015 & 0.016 & 0.025 \\
    \cmidrule(lr){2-10}
    & \multirow{4}{*}{$70\%$} & \multirow{2}{*}{500}
        & gbm        & 0.005 & 0.007 & 0.007 & 0.008 & 0.009 & 0.042 \\
    & & & EBWMA-iEBW & 0.013 & 0.018 & 0.021 & 0.021 & 0.024 & 0.046 \\
    \cmidrule(lr){3-10}
    & & \multirow{2}{*}{1000}
        & gbm        & 0.003 & 0.004 & 0.004 & 0.004 & 0.005 & 0.010 \\
    & & & EBWMA-iEBW & 0.009 & 0.012 & 0.013 & 0.013 & 0.014 & 0.025 \\
    \midrule
    \multirow{12}{*}{DGP 2} & \multirow{4}{*}{$30\%$} & \multirow{2}{*}{500}
        & gbm        & 0.022 & 0.045 & 0.056 & 0.061 & 0.070 & 0.213 \\
    & & & EBWMA-iEBW & 0.048 & 0.071 & 0.081 & 0.083 & 0.092 & 0.173 \\
    \cmidrule(lr){3-10}
    & & \multirow{2}{*}{1000}
        & gbm        & 0.016 & 0.030 & 0.036 & 0.039 & 0.045 & 0.154 \\
    & & & EBWMA-iEBW & 0.035 & 0.058 & 0.067 & 0.069 & 0.078 & 0.136 \\
    \cmidrule(lr){2-10}
    & \multirow{4}{*}{$50\%$} & \multirow{2}{*}{500}
        & gbm        & 0.012 & 0.022 & 0.028 & 0.030 & 0.034 & 0.115 \\
    & & & EBWMA-iEBW & 0.035 & 0.053 & 0.062 & 0.064 & 0.073 & 0.142 \\
    \cmidrule(lr){3-10}
    & & \multirow{2}{*}{1000}
        & gbm        & 0.009 & 0.016 & 0.020 & 0.022 & 0.027 & 0.077 \\
    & & & EBWMA-iEBW & 0.029 & 0.048 & 0.056 & 0.057 & 0.065 & 0.117 \\
    \cmidrule(lr){2-10}
    & \multirow{4}{*}{$70\%$} & \multirow{2}{*}{500}
        & gbm        & 0.008 & 0.014 & 0.019 & 0.022 & 0.027 & 0.094 \\
    & & & EBWMA-iEBW & 0.031 & 0.049 & 0.057 & 0.060 & 0.068 & 0.120 \\
    \cmidrule(lr){3-10}
    & & \multirow{2}{*}{1000}
        & gbm        & 0.005 & 0.011 & 0.015 & 0.017 & 0.021 & 0.117 \\
    & & & EBWMA-iEBW & 0.024 & 0.044 & 0.052 & 0.054 & 0.061 & 0.111 \\
    \bottomrule
\end{tabular}
\vspace{1mm}
\begin{flushleft}
\footnotesize
Note: The relative maximum weight is $\max_{i\in\mathcal{I}_1} w_i^{1,0} / \sum_{i\in\mathcal{I}_1} w_i^{1,0}$, so that a value of $1/n_1$ corresponds to uniform weights. Summary statistics are taken over the 2{,}000 replicates of each condition.
DGP: data-generating process; gbm: gradient boosting model;
EBWMA-iEBW: energy balancing weights for mediation analysis (iEBW for first stage);
1st Qu: first quartile; 3rd Qu: third quartile.
\end{flushleft}
\end{table}

\begin{table}[htbp]
\centering
\caption{Coefficient of variation of $w^{1,0}$ across data-generating processes and treatment prevalences.}
\label{tbl:supp_sim_results_cv}
\vspace{2mm}
\small
\begin{tabular}{ cccccccccc }
    \toprule
    DGP & $\Pr(A=1)$ & $n$ & Method & Min & 1st Qu & Median & Mean & 3rd Qu & Max \\
    \midrule
    \multirow{12}{*}{DGP 1} & \multirow{4}{*}{$30\%$} & \multirow{2}{*}{500}
        & gbm        & 0.372 & 0.544 & 0.604 & 0.614 & 0.674 & 1.030 \\
    & & & EBWMA-iEBW & 0.795 & 1.015 & 1.075 & 1.085 & 1.148 & 1.501 \\
    \cmidrule(lr){3-10}
    & & \multirow{2}{*}{1000}
        & gbm        & 0.413 & 0.584 & 0.638 & 0.643 & 0.694 & 1.123 \\
    & & & EBWMA-iEBW & 1.018 & 1.180 & 1.230 & 1.239 & 1.292 & 1.594 \\
    \cmidrule(lr){2-10}
    & \multirow{4}{*}{$50\%$} & \multirow{2}{*}{500}
        & gbm        & 0.236 & 0.355 & 0.392 & 0.398 & 0.435 & 0.630 \\
    & & & EBWMA-iEBW & 0.890 & 1.064 & 1.110 & 1.114 & 1.162 & 1.387 \\
    \cmidrule(lr){3-10}
    & & \multirow{2}{*}{1000}
        & gbm        & 0.300 & 0.385 & 0.414 & 0.417 & 0.448 & 0.618 \\
    & & & EBWMA-iEBW & 1.072 & 1.209 & 1.249 & 1.249 & 1.289 & 1.485 \\
    \cmidrule(lr){2-10}
    & \multirow{4}{*}{$70\%$} & \multirow{2}{*}{500}
        & gbm        & 0.145 & 0.254 & 0.286 & 0.292 & 0.323 & 1.001 \\
    & & & EBWMA-iEBW & 1.043 & 1.245 & 1.310 & 1.317 & 1.380 & 1.824 \\
    \cmidrule(lr){3-10}
    & & \multirow{2}{*}{1000}
        & gbm        & 0.170 & 0.265 & 0.292 & 0.297 & 0.323 & 0.567 \\
    & & & EBWMA-iEBW & 1.171 & 1.374 & 1.423 & 1.429 & 1.482 & 1.759 \\
    \midrule
    \multirow{12}{*}{DGP 2} & \multirow{4}{*}{$30\%$} & \multirow{2}{*}{500}
        & gbm        & 0.936 & 1.256 & 1.380 & 1.415 & 1.538 & 2.879 \\
    & & & EBWMA-iEBW & 1.691 & 2.039 & 2.147 & 2.155 & 2.257 & 2.763 \\
    \cmidrule(lr){3-10}
    & & \multirow{2}{*}{1000}
        & gbm        & 1.083 & 1.378 & 1.485 & 1.505 & 1.610 & 2.818 \\
    & & & EBWMA-iEBW & 2.184 & 2.566 & 2.679 & 2.687 & 2.800 & 3.347 \\
    \cmidrule(lr){2-10}
    & \multirow{4}{*}{$50\%$} & \multirow{2}{*}{500}
        & gbm        & 0.725 & 1.018 & 1.095 & 1.112 & 1.185 & 2.012 \\
    & & & EBWMA-iEBW & 1.848 & 2.174 & 2.271 & 2.280 & 2.375 & 2.929 \\
    \cmidrule(lr){3-10}
    & & \multirow{2}{*}{1000}
        & gbm        & 0.894 & 1.132 & 1.204 & 1.227 & 1.303 & 2.032 \\
    & & & EBWMA-iEBW & 2.351 & 2.755 & 2.868 & 2.876 & 2.991 & 3.618 \\
    \cmidrule(lr){2-10}
    & \multirow{4}{*}{$70\%$} & \multirow{2}{*}{500}
        & gbm        & 0.564 & 0.844 & 0.918 & 0.942 & 1.013 & 1.965 \\
    & & & EBWMA-iEBW & 1.919 & 2.356 & 2.471 & 2.481 & 2.598 & 3.207 \\
    \cmidrule(lr){3-10}
    & & \multirow{2}{*}{1000}
        & gbm        & 0.713 & 0.950 & 1.026 & 1.060 & 1.137 & 3.225 \\
    & & & EBWMA-iEBW & 2.572 & 3.009 & 3.141 & 3.156 & 3.295 & 4.017 \\
    \bottomrule
\end{tabular}
\vspace{1mm}
\begin{flushleft}
\footnotesize
Note: Summary statistics are taken over the 2{,}000 replicates of each condition.
DGP: data-generating process; gbm: gradient boosting model;
EBWMA-iEBW: energy balancing weights for mediation analysis (iEBW for first stage);
1st Qu: first quartile; 3rd Qu: third quartile.
\end{flushleft}
\end{table}

\begin{table}[htbp]
\centering
\caption{Optimization convergence proportions for the two-step minimal weights method across data-generating processes and treatment prevalences.}
\label{tbl:sim_results_conv}
\vspace{2mm}
\small
\begin{tabular}{ cccc }
    \toprule
    DGP & $\Pr(A=1)$ & $n$ & Convergence proportion \\
    \midrule
    \multirow{6}{*}{DGP 1} & \multirow{2}{*}{$30\%$}
      &  500 & 1.000 \\
    & & 1000 & 1.000 \\
    \cmidrule(lr){2-4}
    & \multirow{2}{*}{$50\%$}
      &  500 & 1.000 \\
    & & 1000 & 1.000 \\
    \cmidrule(lr){2-4}
    & \multirow{2}{*}{$70\%$}
      &  500 & 1.000 \\
    & & 1000 & 1.000 \\
    \midrule
    \multirow{6}{*}{DGP 2} & \multirow{2}{*}{$30\%$}
      &  500 & 0.474 \\
    & & 1000 & 0.900 \\
    \cmidrule(lr){2-4}
    & \multirow{2}{*}{$50\%$}
      &  500 & 0.950 \\
    & & 1000 & 1.000 \\
    \cmidrule(lr){2-4}
    & \multirow{2}{*}{$70\%$}
      &  500 & 0.991 \\
    & & 1000 & 1.000 \\
    \bottomrule
\end{tabular}
\vspace{1mm}
\begin{flushleft}
\footnotesize
Note: Entries give the proportion of the 2{,}000 replicates in which the optimization converged for all three weight vectors $w^{1,1}$, $w^{0,0}$, and $w^{1,0}$. All metrics reported for 2-step MW are computed over converged replicates only.
DGP: data-generating process.
\end{flushleft}
\end{table}

\begin{table}[htbp]
\centering
\caption{RMSE of TE, NDE, and NIE under outcome model 2, DGP 1.}
\label{tbl:rmse_Y2_dgp1}
\vspace{2mm}
\small
\begin{tabular*}{\linewidth}{@{\extracolsep{\fill}} ccl rrr @{}}
    \toprule
    $n$ & $\Pr(A=1)$ & Method & TE & NDE & NIE \\
    \midrule
    \multirow{12}{*}{500}
      & \multirow{4}{*}{$30\%$}
      & glm        & 7010.061 & 6675.375 & 1587.289 \\
    & & gbm        &  231.297 &  199.093 &  128.107 \\
    & & 2-step MW  &  333.254 &  345.734 &  217.770 \\
    & & EBWMA-iEBW &  168.112 &  153.599 &  170.047 \\
    \cmidrule(lr){2-6}
    & \multirow{4}{*}{$50\%$}
      & glm        & 6263.628 & 5970.663 & 1638.248 \\
    & & gbm        &  189.862 &  174.795 &   95.201 \\
    & & 2-step MW  &  335.636 &  331.037 &  127.701 \\
    & & EBWMA-iEBW &  129.805 &  144.225 &  109.642 \\
    \cmidrule(lr){2-6}
    & \multirow{4}{*}{$70\%$}
      & glm        & 5409.238 & 5172.592 & 1121.577 \\
    & & gbm        &  197.043 &  189.915 &   79.496 \\
    & & 2-step MW  &  423.075 &  417.808 &  140.786 \\
    & & EBWMA-iEBW &  176.198 &  149.739 &   88.365 \\
    \midrule
    \multirow{12}{*}{1000}
      & \multirow{4}{*}{$30\%$}
      & glm        & 7911.022 & 7349.906 & 2109.148 \\
    & & gbm        &  181.615 &  143.182 &  133.413 \\
    & & 2-step MW  &  256.732 &  281.789 &  156.964 \\
    & & EBWMA-iEBW &  125.943 &  115.151 &  160.518 \\
    \cmidrule(lr){2-6}
    & \multirow{4}{*}{$50\%$}
      & glm        & 7070.423 & 6553.285 & 2112.447 \\
    & & gbm        &  130.173 &  122.256 &   94.012 \\
    & & 2-step MW  &  291.250 &  297.405 &   89.400 \\
    & & EBWMA-iEBW &   82.692 &  103.576 &   91.494 \\
    \cmidrule(lr){2-6}
    & \multirow{4}{*}{$70\%$}
      & glm        & 6123.490 & 5844.846 & 1412.858 \\
    & & gbm        &  139.311 &  138.024 &   70.262 \\
    & & 2-step MW  &  377.080 &  381.816 &  101.178 \\
    & & EBWMA-iEBW &  127.345 &  106.573 &   67.095 \\
    \bottomrule
\end{tabular*}
\vspace{1mm}
\begin{flushleft}
\footnotesize
Note: The true effects under this specification are TE $=10.0$, NDE $=10.0$, and NIE $=0.0$ (theoretical values). The true NIE is zero under this specification, so the reported NIE RMSE measures dispersion around a null effect. RMSE values are not comparable with those in Table~\ref{tbl:sim_results_rmse} or Table~\ref{tbl:rmse_Y3_dgp1}, because the scale of $Y$ differs across outcome models.
TE: total effect; NDE: natural direct effect; NIE: natural indirect effect;
glm: generalized linear model; gbm: gradient boosting model; 2-step MW: two-step minimal weights;
EBWMA-iEBW: energy balancing weights for mediation analysis (iEBW for first stage).
\end{flushleft}
\end{table}

\begin{table}[htbp]
\centering
\caption{RMSE of TE, NDE, and NIE under outcome model 2, DGP 2.}
\label{tbl:rmse_Y2_dgp2}
\vspace{2mm}
\small
\begin{tabular*}{\linewidth}{@{\extracolsep{\fill}} ccl rrr @{}}
    \toprule
    $n$ & $\Pr(A=1)$ & Method & TE & NDE & NIE \\
    \midrule
    \multirow{12}{*}{500}
      & \multirow{4}{*}{$30\%$}
      & glm        & 42.156 & 44.747 & 15.965 \\
    & & gbm        &  5.805 &  5.984 &  6.280 \\
    & & 2-step MW  & 10.345 & 18.494 & 21.390 \\
    & & EBWMA-iEBW &  5.316 &  5.904 &  6.127 \\
    \cmidrule(lr){2-6}
    & \multirow{4}{*}{$50\%$}
      & glm        & 55.070 & 55.829 &  9.678 \\
    & & gbm        &  5.046 &  8.389 &  7.567 \\
    & & 2-step MW  &  8.157 & 15.614 & 19.076 \\
    & & EBWMA-iEBW &  4.116 &  5.437 &  6.028 \\
    \cmidrule(lr){2-6}
    & \multirow{4}{*}{$70\%$}
      & glm        & 27.182 & 28.791 &  9.163 \\
    & & gbm        &  5.780 & 11.544 &  9.798 \\
    & & 2-step MW  &  5.982 & 15.638 & 16.573 \\
    & & EBWMA-iEBW &  4.222 &  5.184 &  6.020 \\
    \midrule
    \multirow{12}{*}{1000}
      & \multirow{4}{*}{$30\%$}
      & glm        & 28.015 & 34.900 & 19.039 \\
    & & gbm        &  3.777 &  4.445 &  5.116 \\
    & & 2-step MW  &  9.793 & 15.037 & 19.329 \\
    & & EBWMA-iEBW &  3.635 &  4.477 &  4.905 \\
    \cmidrule(lr){2-6}
    & \multirow{4}{*}{$50\%$}
      & glm        & 15.839 & 21.943 & 14.788 \\
    & & gbm        &  3.377 &  6.962 &  6.684 \\
    & & 2-step MW  &  7.765 & 13.944 & 18.358 \\
    & & EBWMA-iEBW &  3.034 &  4.171 &  4.600 \\
    \cmidrule(lr){2-6}
    & \multirow{4}{*}{$70\%$}
      & glm        & 19.614 & 21.818 &  9.110 \\
    & & gbm        &  3.770 & 10.259 &  9.143 \\
    & & 2-step MW  &  5.028 & 13.423 & 15.394 \\
    & & EBWMA-iEBW &  3.256 &  3.858 &  4.633 \\
    \bottomrule
\end{tabular*}
\vspace{1mm}
\begin{flushleft}
\footnotesize
Note: The true effects under this specification are TE $=25.0$, NDE $=5.0$, and NIE $=20.0$ (theoretical values). RMSE values are not comparable with those in Table~\ref{tbl:sim_results_rmse} or Table~\ref{tbl:rmse_Y3_dgp2}, because the scale of $Y$ differs across outcome models.
TE: total effect; NDE: natural direct effect; NIE: natural indirect effect;
glm: generalized linear model; gbm: gradient boosting model; 2-step MW: two-step minimal weights;
EBWMA-iEBW: energy balancing weights for mediation analysis (iEBW for first stage).
\end{flushleft}
\end{table}

\begin{table}[htbp]
\centering
\caption{RMSE of TE, NDE, and NIE under outcome model 3, DGP 1.}
\label{tbl:rmse_Y3_dgp1}
\vspace{2mm}
\small
\begin{tabular*}{\linewidth}{@{\extracolsep{\fill}} ccl rrr @{}}
    \toprule
    $n$ & $\Pr(A=1)$ & Method & TE & NDE & NIE \\
    \midrule
    \multirow{12}{*}{500}
      & \multirow{4}{*}{$30\%$}
      & glm        & 0.942 & 0.858 & 0.442 \\
    & & gbm        & 0.220 & 0.180 & 0.068 \\
    & & 2-step MW  & 0.174 & 0.143 & 0.106 \\
    & & EBWMA-iEBW & 0.140 & 0.149 & 0.087 \\
    \cmidrule(lr){2-6}
    & \multirow{4}{*}{$50\%$}
      & glm        & 0.869 & 0.780 & 0.332 \\
    & & gbm        & 0.210 & 0.179 & 0.049 \\
    & & 2-step MW  & 0.179 & 0.119 & 0.105 \\
    & & EBWMA-iEBW & 0.122 & 0.145 & 0.099 \\
    \cmidrule(lr){2-6}
    & \multirow{4}{*}{$70\%$}
      & glm        & 0.802 & 0.738 & 0.180 \\
    & & gbm        & 0.224 & 0.193 & 0.053 \\
    & & 2-step MW  & 0.196 & 0.137 & 0.099 \\
    & & EBWMA-iEBW & 0.140 & 0.166 & 0.110 \\
    \midrule
    \multirow{12}{*}{1000}
      & \multirow{4}{*}{$30\%$}
      & glm        & 0.954 & 0.842 & 0.488 \\
    & & gbm        & 0.174 & 0.140 & 0.054 \\
    & & 2-step MW  & 0.153 & 0.100 & 0.099 \\
    & & EBWMA-iEBW & 0.100 & 0.122 & 0.095 \\
    \cmidrule(lr){2-6}
    & \multirow{4}{*}{$50\%$}
      & glm        & 0.887 & 0.763 & 0.346 \\
    & & gbm        & 0.168 & 0.146 & 0.037 \\
    & & 2-step MW  & 0.171 & 0.096 & 0.102 \\
    & & EBWMA-iEBW & 0.088 & 0.126 & 0.106 \\
    \cmidrule(lr){2-6}
    & \multirow{4}{*}{$70\%$}
      & glm        & 0.799 & 0.722 & 0.202 \\
    & & gbm        & 0.180 & 0.156 & 0.041 \\
    & & 2-step MW  & 0.185 & 0.111 & 0.098 \\
    & & EBWMA-iEBW & 0.102 & 0.141 & 0.116 \\
    \bottomrule
\end{tabular*}
\vspace{1mm}
\begin{flushleft}
\footnotesize
Note: The true effects under this specification are TE $=10.1$, NDE $=10.0$, and NIE $=0.1$ (calculated through a Monte Carlo simulation). The true NIE is small on this scale, so the RMSE values for the NIE are large relative to the estimand. RMSE values are not comparable with those in Table~\ref{tbl:sim_results_rmse} or Table~\ref{tbl:rmse_Y2_dgp1}, because the scale of $Y$ differs across outcome models.
TE: total effect; NDE: natural direct effect; NIE: natural indirect effect;
glm: generalized linear model; gbm: gradient boosting model; 2-step MW: two-step minimal weights;
EBWMA-iEBW: energy balancing weights for mediation analysis (iEBW for first stage).
\end{flushleft}
\end{table}

\begin{table}[htbp]
\centering
\caption{RMSE of TE, NDE, and NIE under outcome model 3, DGP 2.}
\label{tbl:rmse_Y3_dgp2}
\vspace{2mm}
\small
\begin{tabular*}{\linewidth}{@{\extracolsep{\fill}} ccl rrr @{}}
    \toprule
    $n$ & $\Pr(A=1)$ & Method & TE & NDE & NIE \\
    \midrule
    \multirow{12}{*}{500}
      & \multirow{4}{*}{$30\%$}
      & glm        & 10.757 & 12.657 & 7.169 \\
    & & gbm        &  1.252 &  1.860 & 1.469 \\
    & & 2-step MW  &  2.305 &  3.418 & 3.888 \\
    & & EBWMA-iEBW &  0.721 &  1.293 & 0.876 \\
    \cmidrule(lr){2-6}
    & \multirow{4}{*}{$50\%$}
      & glm        & 12.116 & 12.651 & 4.428 \\
    & & gbm        &  1.227 &  1.943 & 1.215 \\
    & & 2-step MW  &  1.545 &  3.105 & 3.584 \\
    & & EBWMA-iEBW &  0.544 &  0.831 & 0.634 \\
    \cmidrule(lr){2-6}
    & \multirow{4}{*}{$70\%$}
      & glm        &  8.467 &  9.217 & 4.470 \\
    & & gbm        &  1.788 &  2.424 & 1.126 \\
    & & 2-step MW  &  1.667 &  3.749 & 3.067 \\
    & & EBWMA-iEBW &  0.739 &  0.698 & 0.655 \\
    \midrule
    \multirow{12}{*}{1000}
      & \multirow{4}{*}{$30\%$}
      & glm        &  8.010 & 10.843 & 8.110 \\
    & & gbm        &  0.815 &  1.342 & 1.181 \\
    & & 2-step MW  &  2.344 &  2.734 & 3.627 \\
    & & EBWMA-iEBW &  0.472 &  0.942 & 0.686 \\
    \cmidrule(lr){2-6}
    & \multirow{4}{*}{$50\%$}
      & glm        &  5.878 &  7.826 & 6.376 \\
    & & gbm        &  0.819 &  1.643 & 1.144 \\
    & & 2-step MW  &  1.384 &  2.736 & 3.549 \\
    & & EBWMA-iEBW &  0.353 &  0.598 & 0.488 \\
    \cmidrule(lr){2-6}
    & \multirow{4}{*}{$70\%$}
      & glm        &  6.469 &  6.612 & 3.691 \\
    & & gbm        &  1.376 &  2.304 & 1.201 \\
    & & 2-step MW  &  1.172 &  3.440 & 2.891 \\
    & & EBWMA-iEBW &  0.456 &  0.493 & 0.479 \\
    \bottomrule
\end{tabular*}
\vspace{1mm}
\begin{flushleft}
\footnotesize
Note: The true effects under this specification are TE $=5.5$, NDE $=5.0$, and NIE $=0.5$ (calculated through a Monte Carlo simulation). RMSE values are not comparable with those in Table~\ref{tbl:sim_results_rmse} or Table~\ref{tbl:rmse_Y2_dgp2}, because the scale of $Y$ differs across outcome models.
TE: total effect; NDE: natural direct effect; NIE: natural indirect effect;
glm: generalized linear model; gbm: gradient boosting model; 2-step MW: two-step minimal weights;
EBWMA-iEBW: energy balancing weights for mediation analysis (iEBW for first stage).
\end{flushleft}
\end{table}

\begin{table}[htbp]
\centering
\caption{Standardized mean differences of covariates and the mediator in the NHEFS analysis.}
\label{tbl:supp_realdata_smd}
\vspace{2mm}
\small
\begin{tabular*}{\linewidth}{@{\extracolsep{\fill}} l rrrrr @{}}
    \toprule
    Variable & Unadjusted & glm & gbm & 2-step MW & EBWMA-iEBW \\
    \midrule
    \multicolumn{6}{l}{\textit{Panel A: $w^{1,1}$ versus $w^{0,0}$}}\\
    \midrule
    Age                          & 0.24067 & 0.28720 & 0.02515 & 0.06321 & 0.00124 \\
    Sex                          & 0.05748 & 0.07940 & 0.00805 & 0.06575 & 0.00009 \\
    Race                         & 0.28205 & 0.47616 & 0.06197 & 0.10075 & 0.00622 \\
    Physical activity (moderate) & 0.07457 & 0.01785 & 0.00729 & 0.06248 & 0.00072 \\
    Physical activity (inactive) & 0.05576 & 0.06533 & 0.03495 & 0.05488 & 0.00254 \\
    Weight in 1982               & 0.04585 & ---     & ---     & ---     & ---     \\
    \midrule
    \multicolumn{6}{l}{\textit{Panel B: $w^{1,0}$ versus $w^{0,0}$}}\\
    \midrule
    Age                          & 0.24067 & 0.28911 & 0.03516 & 0.01405 & 0.00191 \\
    Sex                          & 0.05748 & 0.07547 & 0.01894 & 0.05012 & 0.01080 \\
    Race                         & 0.28205 & 0.47306 & 0.09428 & 0.02402 & 0.02821 \\
    Physical activity (moderate) & 0.07457 & 0.01791 & 0.00819 & 0.05326 & 0.00316 \\
    Physical activity (inactive) & 0.05576 & 0.06576 & 0.01737 & 0.04935 & 0.00399 \\
    Weight in 1982               & 0.04585 & 0.01906 & 0.03988 & 0.03436 & 0.00009 \\
    \bottomrule
\end{tabular*}
\vspace{1mm}
\begin{flushleft}
\footnotesize
Note: Standardized mean differences are computed against the pooled standard deviation before weighting. The mediator does not enter the first-stage comparison in Panel A and is therefore not reported there. Standardized mean differences compare first moments only.
2-step MW results use $\delta_k=0.05$.
glm: generalized linear model; gbm: gradient boosting model; 2-step MW: two-step minimal weights;
EBWMA-iEBW: energy balancing weights for mediation analysis (iEBW for first stage).
\end{flushleft}
\end{table}

\end{document}